\documentclass[twocolumn,showpacs,preprintnumbers,amsmath,amssymb]{revtex4}

\usepackage{epsf}
\usepackage{rotating}
\usepackage{color}

\usepackage{graphicx,epsfig}
\usepackage{dcolumn}
\usepackage{bm}

\def\cH{{\cal H}}

\def\cC{{\cal C}}
\def\cZ{{\cal Z}}

\newcommand{\req}[1]{Eq.~(\ref{#1})}
\newcommand{\avg}[1]{\langle #1\rangle}

\newcommand{\fig}[1]{Fig.~\ref{#1}}

\DeclareMathOperator*{\trace}{{\mathbf{Tr}}}		

\newcommand{\size}{r}

\newcommand{\hin}{{u}}
\newcommand{\hout}{{v}}
\newcommand{\qab}{{q_{\alpha\beta}}}
\newcommand{\ma}{{m_{\alpha}}}
\newcommand{\sia}{{s_{i\alpha}}}
\newcommand{\sib}{{s_{i\beta}}}

\newcommand{\sgia}{{\sigma_{i\alpha}}}
\newcommand{\sgib}{{\sigma_{i\beta}}}

\newcommand{\isingZ}{\cZ_{\rm Ising}}

\newcommand{\isingS}{S_{\rm Ising}}

\newcommand{\skZ}{\cZ_{\rm SK}}
\newcommand{\skXi}{\Xi_{\rm SK}}
\newcommand{\skS}{S_{\rm SK}}
\newcommand{\skSize}{\size_{\rm SK}}

\newcommand{\msm}{m_{s\sigma}}

\newcommand{\qssmm}{q_{s\sigma s\sigma}}
\newcommand{\mm}{m_{\sigma}}

\newcommand{\cut}[1]{{}}
\newcommand{\etal}[1]{\emph{~et al.}}

\newcommand{\bchange}[1]{{}}

\begin{document}

\preprint{}

\title[Title]
{Self-sustained Clusters and Ergodicity Breaking in Spin Models}
\author{Chi Ho Yeung and David Saad}
\affiliation{Nonlinearity and Complexity Research Group, Aston University, Birmingham B4 7ET, United Kingdom}

\date{\today}

\begin{abstract}
The emergence of self-sustained clusters and their role in ergodicity breaking is investigated in fully connected Ising and Sherrington-Kirkpatick (SK) models.  The analysis reveals a clustering behavior at various parameter regimes, as well as yet unobserved phenomena such as the absence of non-trivial clusters in the Ising ferromagnetic and paramagnetic regimes, the formation of restricted spin clusters in SK spin glass and a first order phase transition in cluster sizes in the SK ferromagnet. The method could be adapted to investigate other spin models.
\end{abstract}
\pacs{75.10.Nr, 05.20.-y, 89.20.-a}

\maketitle

Spin glasses are magnetic materials characterized by extremely slow magnetization relaxation in the absence of external field~\cite{nordblad86, binder86}. Several models have been developed to explain their behavior~\cite{edwards75, sherrington75} which in turn have revealed a rich physical picture of a rugged free energy landscape~\cite{binder86, mezard87}. Remarkably, the physics of spin glasses has a non-trivial connection to interdisciplinary applications including image processing, error correcting codes, neural networks and combinatorial optimization~\cite{nishimori01, opper01}. Its connection to structural glasses and supercooled liquids have also been explored to explain the physics below the glassy temperature~\cite{kirkpatrick87, karmakara13}.

Among the various spin glass models, the Sherrington-Kirkpatrick (SK) model~\cite{sherrington75} is arguably the most studied. One of the most intriguing features of large scale disordered systems in general and the SK model in particular, is the breaking of ergodicity in some parameter regimes (e.g., temperature, strength of interactions or topology), particularly in the spin-glass phase where it manifests itself through a complex symmetry structure of order parameters that describe macroscopically the corresponding solution space.
Although macroscopic properties of the SK model are relatively clear its microscopic features are less understood~\cite{palmer79}, in particular the existence of stable domains that are independent of the remainder of the system; these are important for gaining insight into the mechanism that gives rise to ergodicity breaking and the physical picture of spin glasses.

In this Letter, we examine analytically the existence of self-sustained spin clusters in fully connected Ising and SK models. We remark that a similar behavior, termed as \emph{backbones} or \emph{frozen variables} in sparse systems~\cite{zhou05, zdeborova07}, is induced by the topological disorder and is therefore somewhat different from the self-sustained clusters studied here; nevertheless, sparse topologies could be analyzed by extending the method presented here. We study the existence and nature of self-sustained clusters in various phases, the dependence of their sizes on system parameters and the existence of phase transitions with respect to cluster sizes. The Ising model will be analyzed first, followed by a more involved analysis of the SK model.

\begin{figure}[th]
\centerline{\epsfig{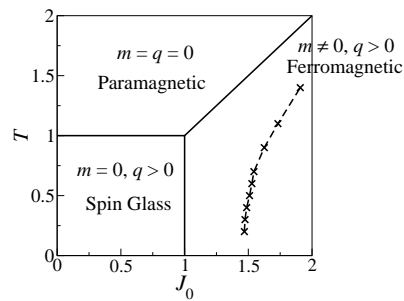}}
\vspace{-0.3cm}
\caption{Phase diagram of the SK model as a function of coupling mean $J_0$ and temperature $T$. Right hand side of the dashed line corresponds to the region where a first order phase transition in cluster sizes is observed, i.e. self-sustained clusters of certain size are absent.}
\label{fig_skphase}
\end{figure}

\emph{\underline{Models}} - The SK model comprises $N$ spin variables, any two of which $i$ and $j$ interact via a ferromagnetic ($J_{ij}>0$) or anti-ferromagnetic ($J_{ij}<0$) symmetric coupling ($J_{ij}\!=\!J_{ji}$). Coupling variables are randomly drawn from a Gaussian distribution of mean $J_0/N$ and variance $J^2/N$; the corresponding Hamiltonian is given by $\cH_{\rm SK}\!=\!-\sum_{(ij)}J_{ij}s_i s_j$, which sums over all un-ordered spin pairs $(i\neq j)$. The \emph{infinite-range Ising model} is a special case of the SK model with $J\!=\!0$ or $J_0\!\to\!\infty$, and the corresponding Hamiltonian is $\cH_{\rm Ising}\!=-J_0\!\sum_{(ij)}s_is_j/N$.

To compute physical quantities of interest, one uses the identity $\ln\cZ = \lim_{n\to 0}(\cZ^n-1)/n$ to carry out an averages over quenched variables, replacing the average of $\ln\cZ$ by that of the \emph{replicated partition function} $\cZ^n$~\cite{mezard87}. As $N\to\infty$, solutions space is described by the magnetization and inter-replica spin correlation order parameters
\begin{align}
\ma = \frac{1}{N}\sum_i \sia,\qquad
\qab = \frac{1}{N}\sum_i \sia\sib,
\end{align}
where $\alpha,\beta\!=\!1,\ldots,n$ are replica indices. An ultrametric structure of the order parameter symmetry is then used to facilitate the calculation, the simplest of which is the \emph{replica-symmetric} (RS) ansatz, where one substitutes $\ma\!=\!m$ for all $\alpha$ and $\qab\!=\!q$ for all $\alpha\!\neq\!\beta$. The various phases observed in the model are expressed by the values of $m$ and $q$, for instance the paramagnetic  $(m\!=\!q\!=\!0)$, ferromagnetic $(m\!\neq\! 0, q\!>\!0)$ and spin glass phases $(m\!=\!0, q\!>\!0)$ as shown in \fig{fig_skphase}.

\emph{\underline{Self-sustained clusters-}} Denote a set $\cC$ of spin variables; for each spin $i\in \cC$ we define \emph{in-cluster} and \emph{out-cluster} magnetic fields $\hin_i = \sum_{j\in \cC}J_{ij}s_j$ and $\hout_i = \sum_{j\notin \cC}J_{ij}s_j$ induced by spins \emph{in} and \emph{out} of $\cC$, respectively. The total magnetic field experienced by spin $i$ is $h_i = u_i + v_i$. The set $\cC$ is \emph{self-sustained} if
\begin{align}
\label{eq_cond}
|\hin_i| > |\hout_i|, \qquad \forall i\in \cC.
\end{align}
In other words, the magnetic field experienced by each individual spin $i$ in $\cC$ is dominated by the contributions of peer spin variables in $\cC$. We remark that our framework can accommodate other cluster definitions.

To obtain the distribution of clusters, we denote $\Omega(\size)$ to be the number of self-sustained clusters of normalized size $\size=|\cC|/N$. Since $\ln\Omega(\size)$ is an extensive quantity, we define the entropy of clusters to be $S(\size)\!=\![\ln\Omega(\size)]/N$. For instance, one can easily compute $S(\size)$ of the Ising model at zero temperature $T\!=\!0$ where all spins are aligned. Since the couplings are uniform, \req{eq_cond} is satisfied for a set $\cC$ if $\size\!>\!0.5$. Indeed, any grouping with at least half of the spins is self-sustained, which implies $\Omega(\size)\!=\!C^N_{Nr}\!=\!N!/[(\size N)!(N-\size N)!]$ and $S(\size)\!=\!-\size\ln\size\!-\!(1\!-\!\size)\ln(1\!-\!\size)$ for $\size\!>\!0.5$; and $\Omega(\size)\!=\!0$ and $S(\size)\!=\!-\infty$ otherwise. We note that using this definition, self-sustained clusters which are subsets of larger self-sustained clusters are also counted.

We further define a variable $\sigma_i=1, -1$ to identify cases when spin $i$ is included in or excluded from the cluster, respectively. Thus, the cluster size $\size\!=\!\sum_i (1\!+\!\sigma_i)/2$. One can then define an \emph{indicator function}
\begin{align}
	 w\left(\{\sigma_i\},\!\{s_i\},\!\{J_{ij}\}\right)\!=\!\prod_{i}\!\left[\frac{1\!-\!\sigma_i}{2}+\frac{1\!+\!\sigma_i}{2}\Theta\!\left(u_i^2\!-\!v_i^2\right)\right],
\end{align}
where the step function $\Theta(x) = 0, 1$ for the cases $x\!<\!0$ and $x\!>\!0$, respectively. It turns out that the value of $\Theta(0)$ is crucial in the paramagnetic phase as will be discussed later. Thus, $w\!=\!1$ if the cluster defined by the set $\{\sigma_i\!=\!1\}$ is self-sustained, and $w=0$ otherwise.

\emph{\underline{Ising Model}} -
To derive $S(\size)$ for the fully connected Ising model at any temperature $T$ we uniformly sample spin configurations of given magnetization $m\!=\!\sum_i s_i/N$, as it uniquely defines the model's macroscopic properties. It is sufficient to introduce an \emph{operator partition function} which measures the entropy $S(\size)$ of clusters given $m$:
\begin{align}
\label{eq_isingZ}
&\isingZ(\gamma, m)
\\
&\qquad
\!=\! \trace_{\{s_i\}}\trace_{\{\sigma_i\}}w(\{\sigma_i\}, \{s_i\})\delta\left(\frac{\sum_i\!s_i}{N} - m\right)e^{\gamma\frac{\sum_i (1\!+\!\sigma_i)}{2}}
\nonumber
\end{align}
where the dependence of $w$ on $\{J_{ij}\}$ is omitted as they are all identical ($J_0$). The parameter $\gamma$ plays the role of \emph{pesudo-temperature} conjugate to the cluster size $\sum_i(1\!+\!\sigma_i)/2$; by computing $\cZ$, one obtains the entropy $S(\gamma)$ and cluster size $\size(\gamma)$ as a function of $\gamma$ leading to $S(\size)$.

Details of the calculation are found in the \emph{Supplementary Information} (SI); here we briefly describe the solution. In the limit of $N\to\infty$, $\isingZ$ is given by
\begin{align}
\label{eq_isingZ_solution}
	\isingZ(\gamma, m)\!=\!A(m)[1+\Theta( \msm m)e^\gamma]^N
\end{align}
where the variable $\msm=\sum_i s_i \sigma_i/N$ and its value is given self-consistently by the equation
\begin{align}
\label{eq_msm}
	\msm = m\left(\frac{\Theta(\msm m)e^\gamma -1}{\Theta(\msm m)e^\gamma +1}\right).
\end{align}
The prefactor $A(m)$ in \req{eq_isingZ_solution} is given by
\begin{align}
\label{eq_lnA}
	A(m) = e^{-\beta N J_0 m^2}[2\cosh(\beta J_0 m)]^N,
\end{align}
the entropic contribution of spin configurations $\{s_i\}$, i.e., $A(m)\!=\!\trace_{\{s_i\}}\delta(\sum_i s_i/N - m)$. Indeed, the partition function of the Ising model is $e^{-\beta E_{\rm Ising}} A(m)$, with average energy $E_{\rm Ising}=-N J_0 m^2/2$.
\cut{Note that while \req{eq_isingZ} has no explicit dependence on the temperature, it depends on $\beta=1/T$ through $A(m)$ as shown in \req{eq_isingZ_solution}; \req{eq_isingZ_solution} depends on $\beta=1/T$ through $A(m)$; the origin of this dependence is a self-consistent equation satisfying the constraint $\delta(\sum_i s_i/N - m)$ (for details see the \emph{SI}).}

Using \req{eq_isingZ_solution}, one can drive the cluster size $\size(\gamma)$ by
\begin{align}
\label{eq_isingSize}
	\size_{\rm Ising}(\gamma, m)
	= \frac{1}{N}\frac{\partial \ln\isingZ}{\partial\gamma}
	= \frac{\Theta(\msm m)e^\gamma}{\Theta(\msm m)e^\gamma +1}~.
\end{align}
To compute the entropy $S(\gamma)$, one subtracts the entropic contribution $\ln A$ from $\ln\isingZ$ and apply the Legendre transformation to obtain
\begin{align}
\label{eq_isingS}
\isingS(\gamma, m)&=\!\frac{1}{N}\left[-\gamma\frac{\partial\ln\isingZ}{\partial\gamma}\!+\!\ln\isingZ\!-\!\ln A\right]
\\
&=\!-\frac{\gamma~\Theta(\msm m)e^\gamma}{\Theta(\msm m)e^\gamma +1} + \ln[1\!+\!\Theta( \msm m)e^\gamma].
\nonumber
\end{align}

\begin{figure}
\centerline{\epsfig{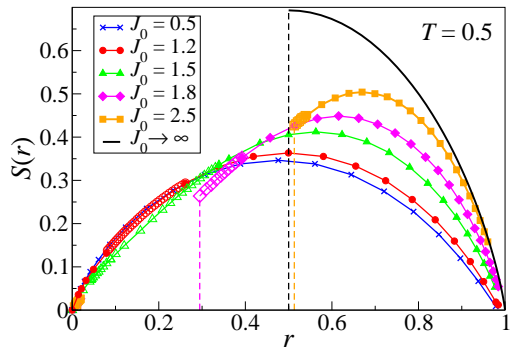}}
\vspace{-0.3cm}
\caption{(Color online) The entropy $S(\size)$ of self-sustained clusters of size $\size$ at $T=0.5$ and various coupling mean values $J_0$ in the SK model.  Curves with open symbols have been obtained with a fine resolution of  $\gamma$ at intervals of $0.02$. The black line $J_0\rightarrow\infty$ corresponds to $S(\size)$ in Ising ferromagnet.}
\label{fig_sk}
\end{figure}

To obtain $S(\size)$, we assume $m\!>\!0$ and solve \req{eq_msm} to obtain $\msm\! =\!-m$ for $\gamma\!<\!0$,
$\msm\! =\!m\left(\frac{e^\gamma-1}{e^\gamma+1}\right)$ for $\gamma\! \ge\! 0$ and no solution in the range $-m\!<\!\msm\!<\!0$; as a result
\begin{align}
\label{eq_isingSr}
S(\size) =
\begin{cases}
0 & \size=0
\\
-\infty & 0<\size<0.5
\\
-\size\ln \size - (1-\size)\ln(1-\size) & \size\ge 0.5
\end{cases},
\end{align}
shown by the black line in \fig{fig_sk}. This result is valid for the entire ferromagnetic phase ($m\!\neq\!0$) and is consistent with $S(r)$ at $T=0$ obtained by simple counting. It implies that in the ferromagnetic phase, regardless of $T$ and $m$, clusters that include at least half of the spins are self-sustained and the magnetization is uniform over any subset of spins even for small $m$ values. Alternatively, one calculates the \emph{in-cluster} and \emph{out-cluster magnetization},
\begin{align}
\label{eq_inclusterM}
\avg{s_i}_{\sigma_i=1} = \frac{m+\msm}{2r}, \qquad
\avg{s_i}_{\sigma_i=-1} = \frac{m-\msm}{2(1-r)}~,
\end{align}
respectively, to show [using Eqs.~(\ref{eq_msm}) and (\ref{eq_isingSize})] that self-sustained clusters have the same magnetization as the out-cluster spins, $\avg{s_i}_{\sigma_i=1}\!=\!\avg{s_i}_{\sigma_i=-1}\!=\!m$.

For the paramagnetic phase, $S(\size)$ is ambiguous since $m=0$ and $|u_i|=|v_i|=0$ in \req{eq_cond}; it thus depends on the definition of $\Theta(0)$ in Eqs.~(\ref{eq_isingSize})~and~(\ref{eq_isingS}). Only trivial self-sustained clusters are observed: the choice $\Theta(0)=1$ results in $S(\size)\!=\!-\size\ln \size\!-\!(1\!-\!\size)\ln(1\!-\!\size)$ for all cluster sizes $0\le\size\le 1$, implying that any subset of spins is considered self-sustained, while for $\Theta(0)=0$, $S(0)=0$ and $S(\size)=-\infty$, $\forall \size\!\neq\!0$, implying that no self-sustained clusters exist. We note that from \req{eq_inclusterM}, $\avg{s_i}_{\sigma_i=1}\!=\!\avg{s_i}_{\sigma_i=-1}\!=0$ regardless of the value of $\Theta(0)$, which implies that magnetized domains are always absent from the paramagnetic phase.

\emph{\underline{SK Model}} -
Similarly, in the SK model we uniformly draw system configurations from a distribution defined by the order parameters $\{\ma\}$ and $\{\qab\}$, and introduce an operator partition function which measures $S(\size)$ given $\{\ma\}$ and $\{\qab\}$. Unlike the Ising model with a single order parameter $m$, the order parameters in the SK model are labeled by replica indices, we thus define a \emph{replicated} operator partition function
\begin{align}
&\skXi(\gamma,\!\{\ma\},\!\{\qab\},n)
\\
&\!=\!\trace_{\{J_{ij}\}}\trace_{\{\sia\}}\trace_{\{\sgia\}}e^{\gamma\sum_{i, \alpha}\frac{1+\sgia}{2}}\prod_{\alpha}w(\{\sgia\},\!\{\sia\},\!\{J_{ij}\})
\nonumber\\
&\times
\prod_{\alpha}\delta\left(\frac{\sum_i\sia}{N}\!-\!\ma\right)
\prod_{\alpha\beta}\delta\left(\frac{\sum_i\sia\sib}{N}\!-\!\qab\right) P(\mathbf{J}).
\nonumber
\end{align}

We further define the corresponding \emph{un-replicated} partition function with respect to spin configurations as $\skZ[\gamma,\!P(\ma),\!P(\qab)]$, such that $P(\ma)$ and $P(\qab)$ are the distributions of $\ma$ and $\qab$ in the limit $n\to 0$. The logarithm $\ln\skZ$ is given by
\begin{align}
\ln\skZ[\gamma,\!P(\ma),\!P(\qab)] = \lim_{n\to\infty}\frac{\skXi-1}{n}.
\end{align}
To find the \emph{exact} form of $P(\ma)$ and $P(\qab)$ in the spin glass phase requires the \emph{full replica sysmetric breaking} (full-RSB) ansatz, which is in principle feasible but very difficult. We will thus compute $\ln\skZ$ under the \emph{replica sysmetric} (RS) ansatz, where $P(\ma)\!=\!\delta(\ma\!-\!m)$ and $P(\qab)\!=\!\delta(\qab\!-\!q)(1-\delta_{a,b})+\delta(\qab\!-\!1)\delta_{a,b}$ such that $\ln\skZ$ only depends on the variables $\gamma, m$ and $q$.

Even with the RS ansatz, the calculation of $\ln\skZ$ is rather involved. We will thus describe the main rationale and results and refer readers to the \emph{SI} for details. To obtain $S(\size)$, we compute $\skSize(\gamma, m, q)$ and $\skS(\gamma, m, q)$ by similar equations to Eqs.~(\ref{eq_isingSize}) and (\ref{eq_isingS}) with $\ln\isingZ$ replaced by $\ln\skZ$ and $\ln A$ replaced by the spin entropic contribution in the SK model
\begin{align}
&\ln B(m,q) = \lim_{n\to 0}\frac{1}{n}
\\
&\!\times\!\Bigg[\!\trace_{\{\sia\}}\!
\prod_{\alpha}\delta\left(\frac{\sum_i\sia}{N}\!-\!m\right)\!
\prod_{\alpha\neq\beta}\delta\left(\frac{\sum_i\sia\sib}{N}\!-\!q\right)\!-\!1\Bigg].
\nonumber
\end{align}

\begin{figure}
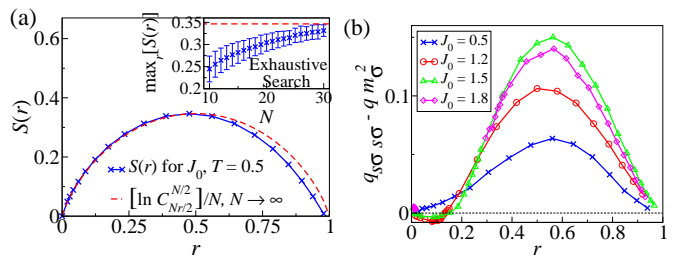

\centerline{\epsfig{figure=sg_entropy3.eps, width=0.5\linewidth}
\epsfig{figure=qssmmDiff.eps, width=0.5\linewidth}}
\vspace{-0.2cm}
\caption{(Color online) (a) Entropy $S(\size)$ of self-sustained clusters in the spin glasse case with $J_0\!=\!0.5, T\!=\!0.5$, compared to the limit of $[\ln\!C^{N/2}_{Nr/2}]/N$ as $N\to\infty$. \emph{Inset}: the maximum of $S(\size)$ obtained by exhaustive search in the ground state of small SK systems with $J_0=0.01$, compared to the value of $S(0.5)\!=\![\ln\!C^{N/2}_{N/4}]/N\!=\!0.346$ (red dashed line); both mean values and standard deviation are shown. (b) The difference $q_{s\sigma s\sigma}-qm_\sigma^2$ as a function of $r$ for various $J_0$ values.}
\label{fig_sg}
\end{figure}

\begin{figure}[t]
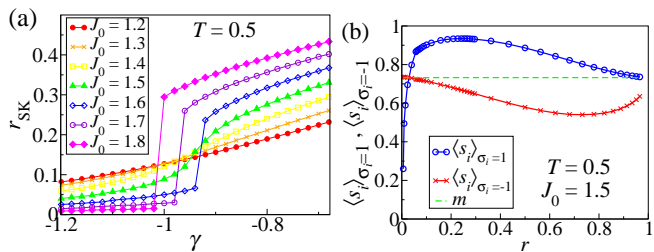

\centerline{\epsfig{figure=size05c.eps, width=0.5\linewidth}
\epsfig{figure=clusterM.eps, width=0.48\linewidth}}
\vspace{-0.2cm}
\caption{(Color online) (a) The self-sustained cluster size $\skSize$ as a function  of $\gamma$ at temperature $T\!=\!0.5$ with $J_0$ from $1.2$ to $1.8$. (b) In-cluster and out-cluster magnetizations as a function of the cluster size $r$.}
\label{fig_size}
\vspace{-0.2cm}
\end{figure}

Figure~\ref{fig_sk} shows $S(r)$ as a function of $J_0$ at $T=0.5$. Remarkably, in the spin glass phase (e.g., $J_0\!=\!0.5$) cluster entropies exhibit a similar general shape to those obtained by counting in a uniform spin configuration but with degrees of freedom reduced (almost exactly) by \emph{half}
\begin{align}
\label{eq_sgSr}
S(r)\!\approx\!\frac{-\!\size\ln\size\!-\! (1-\size)\ln(1-\size)}{2}=\frac{\ln C^{N/2}_{Nr/2}}{N},
\end{align}
as shown in \fig{fig_sg}(a). We observe that this picture holds in the spin glass phase regardless of the values of $T$ and $J_0$. To test the validity of this result we studied numerically $S(r)$ in small SK systems by exhaustive search. The inset of \fig{fig_sg}(a) shows that $\max_r[S(r)]$ is approaching $[\ln C^{N/2}_{N/4}]/N$ as $N$ increases in agreement with the theoretical predictions.

This profile of self-sustained clusters is consistent with our understanding of the spin-glass phase: firstly, it shows a gap between the trivial cluster that encompasses the entire system ($r\!=\!1$) and the exponential number of smaller self-sustained clusters which presumably correspond to suboptimal solutions; secondly, it shows that smaller size self-sustained clusters are determined by approximately half of their constituent spins while the other half are fixed by inherent system correlations.

To further understand the relation between self-sustained clusters and ergodicity breaking, we examine the difference $d=\qssmm - q\mm^2$ where
\begin{align}
\qssmm = [\avg{\sia\sgia\sib\sgib}_{i, \alpha,\beta}], \quad \mm=[\avg{\sgia}_{i, \alpha}],
\end{align}
with $[\dots]$ corresponding to the average over coupling disorders. One expects $d\!=\!0$ when the spin-configuration overlap between two replica is uncorrelated with cluster affiliations; on the other hand, $d\!>\!0$ when correlated spin-configuration in two replica tend to have correlated cluster associations. Figure~\ref{fig_sg}(b) shows $d$ as a function of $r$ for various $J_0$ values. In the spin glass phase ($J_0\!=\!0.5$) $d\!>\!0$ for all $r$, suggesting: (i) the presence of \emph{self-sustained} and \emph {frozen} spin clusters of all size; (ii) not all spin subsets constitute self-sustained clusters, or otherwise $d$ would have vanished. These results suggest that an extensive number of spin flips are required to destabilize or modify self-sustained clusters, which points to the existence of high energy barriers that lead to meta-stable configurations. The same phenomenon is identified as backbone or rigidity in sparse spin systems studied elsewhere. A similar picture emerges in the ferromagnetic phase with small $J_0\!>\!1$, except that $d\!\approx\! 0$ for small $r$ values. This result and the cluster magnetization $\avg{s_i}_{\sigma_i=1}\!\to\!0$ as $r\!\to\!0$ (see \fig{fig_size}(b)) suggest that small self-sustained clusters are not frozen and thus can be easily flipped to merge into larger clusters. All these indicate that ergodicity breaking is most prominent in the spin glass phase.

We continue to examine $S(r)$ by increasing $J_0$, exiting the spin glass to the ferromagnetic phase,
where one expects a different profile of $S(\size)$ than that of \req{eq_sgSr}; this difference is particularly emphasized when one considers the limit of $J_0\to\infty$, which corresponds to \req{eq_isingSr}. The cluster entropy, shown for increasing $J_0$ values in \fig{fig_sk}, exhibits the onset of discontinuity in cluster sizes at $J_0=1.8$, implying the absence of clusters in a range of sizes.  The range where discontinuity occurs increases with $J_0$ until $S(r)$ reduces to \req{eq_isingSr} when $J_0\to\infty$ as shown in the \emph{SI}.
To examine this behavior we plot the expected cluster size $\size$ as a function of $\gamma$ in \fig{fig_size}(a) (higher $\gamma$ selects clusters of a larger size). An abrupt jump in cluster size appears when $J_0\ge 1.6$, resembling a first order transition, which implies the emergence of large and small clusters and the absence of clusters of sizes in between. The phase boundary identifying the onset of this first order phase transition was added to the SK phase diagram in \fig{fig_skphase}, denoted by $\times$ symbols. This phase line marks the emergence of an extensive ferromagnetic domain which grows in size as $J_0$ increases and becomes the trivial cluster in the limit $J_0\!\to\!\infty$.

One should note that $\size(\gamma)$ is not identically zero before the transition point, implying the presence of small clusters in the ferromagnetic phase, which presumably correspond to small spin domains of arbitrary alignment. Figure~\ref{fig_size}(b) shows that the in-cluster magnetization $\avg{s_i}_{\sigma_i=1}\!>\!m$ for the entire range of $r$ except when $r\!\gtrsim\!0$. This result and the out-cluster magnetization $\avg{s_i}_{\sigma_i=-1}\!<\!m$
suggest the presence of local domains of weaker magnetic alignment. We remark that similar magnetization domains do not appear in the Ising ferromagnet, suggesting that coupling disorder is crucial for the formation of such domains.


\emph{\underline{Summary}} -
We showed that self-sustained clusters relate to the formation of meta-stable configurations separated by an extensive number of variables, one of the main features exhibited by disordered systems in the spin-glass phase, which leads to ergodicity breaking.
Such domains have been termed backbone variables elsewhere. We reveal the existence of such clusters in the spin-glass and ferromagnetic phases of the SK model and the absence of non-trivial clusters in the Ising model. Other observations include a first order phase transition in the size of self-sustained clusters and the presence of domains of stronger magnetic alignment in the SK ferromagnetic regime. The role of self-sustained clusters in different spin models, both sparsely and densely connected, is yet to be investigated analytically for gaining insights into the corresponding physical behavior; the new framework and understanding may also play an important role in interdisciplinary applications, particularly the development of optimization algorithms.

This work is supported by the EU project STAMINA (FP7-265496) and Royal Society Grant IE110151.


\end{document}


\title{Supplementary Information
\\ {Self-sustained Clusters and Ergodicity Breaking in Spin Models}}
\author{Chi Ho Yeung and David Saad}

\maketitle

\tableofcontents

\section{Derivation of $\isingZ$}

We would like to study the properties of self-sustained clusters without affecting the Ising system itself. To do that we exploit the fact that the model has been solved previously and is determined macroscopically by the order parameter $m$. We therefore uniformly draw configurations $\mathbf{s}$ that are consistent with the macroscopic description of the model, the value of the parameter $m$.

We start from the operator partition function $\isingZ$ of the Ising model
\begin{align}
\label{eq_isingZ}
&\isingZ(\gamma, M)
\!=\! \trace_{\{s_i\}}\trace_{\{\sigma_i\}}w(\{\sigma_i\}, \{s_i\})\delta\left(\frac{\sum_i\!s_i}{N} - M\right)e^{\gamma\frac{\sum_i (1\!+\!\sigma_i)}{2}},
\end{align}
where we denote the magnetization of the Ising model as capital letter $M$ in this supplementary information, instead of $m$ as in the main paper, to avoid confusion in subsequent derivations. The indicator function $w$ is given by
\begin{align}
\label{eq_w}
	w\left(\{\sigma_i\}, \{s_i\}, \{J_{ij}\}\right)=\prod_{i}\left[\frac{1\!-\!\sigma_i}{2}+\frac{1\!+\!\sigma_i}{2}\Theta\left(u_i^2\!-\!v_i^2\right)\right],
\end{align}
where the step function $\Theta(x) = 0, 1$ for the cases $x<0$ and $x>0$ respectively, and assume $\Theta(0)$ to be either $1$ or $0$, as discussed in the paper; the variable $\sigma_i=-1,1$ corresponds to the case when spin $i$ is included in or excluded from the cluster, respectively. The \emph{in-cluster magnetic field} $\hin_i$ and \emph{out-cluster magnetic field} $\hout_i$ are defined by
\begin{align}
\hin_i = \sum_{j\in \cC}J_{ij}s_j =  \sum_{j}J_{ij}s_j \frac{1+\sigma_j}{2},
\\
\hout_i = \sum_{j\notin \cC}J_{ij}s_j = \sum_{j}J_{ij}s_j \frac{1-\sigma_j}{2}.
\end{align}
Using these definitions, the argument $u_i^2-v_i^2$ in the step function of \req{eq_w} can be simplified as
\begin{align}
u_i^2\!-\!v_i^2 = \left(\sum_{j}J_{ij}s_j \sigma_j\right)\left(\sum_{k}J_{ik}s_k \right).
\end{align}
In order to trace over $\{s_i\}$ and $\{\sigma_i\}$ in \req{eq_isingZ}, one has to factorize the terms in $\isingZ$ over $i$. To achieve the goal, we: (i) use the integral representation of delta function to represent $\delta(\sum_i s_i/N - M)$; (ii) denote $h_i = \sum_{j}J_{ij}s_j$ and $\eta_i=\sum_{j}J_{ij}s_j \sigma_j$. These lead to
\begin{align}
\label{eq_ising_step1}
\isingZ
=& \trace_{\{s_i\}}\trace_{\{\sigma_i\}}\int \frac{d \hat{M}}{2\pi}e^{i M\hat{M} - i\hat{M}\frac{\sum_i\!s_i}{N}}
\nonumber\\
&\times
\prod_i\left[\int \frac{d h_i d\hat{h_i}}{2\pi}\int \frac{d \eta_i d\hat{\eta_i}}{2\pi} e^{i h_i\hat{h}_i -i\hat{h}_i\sum_{j}J_{ij}s_j + i\eta_i\hat{\eta}_i -i\hat{\eta}_i\sum_{j}J_{ij}s_j \sigma_j}\right]
\nonumber\\
&\times \prod_{i}\left[\frac{1\!-\!\sigma_i}{2}+\frac{1\!+\!\sigma_i}{2}\Theta\left(h_i\eta_i\right)\right]
e^{\gamma\frac{\sum_i (1\!+\!\sigma_i)}{2}},
\end{align}
We then substitute $J_{ij}=J_0/N$ in the Ising model case and introduce the following mean-field parameters
\begin{align}
\label{eq_m}
&\ms = \frac{1}{N}\sum_i s_i,
\\
&\msm = \frac{1}{N}\sum_i s_i \sigma_i,
\\
&\mh = -\frac{i}{N}\sum_i \hat{h}_i,
\\
\label{eq_me}
&\me = -\frac{i}{N}\sum_i \hat{\eta}_i.
\end{align}
One should note that $\ms$ is indeed the given magnetization $M$ in \req{eq_isingZ}, so we expect $\ms=M$ to be a consequence of the subsequent  derivation. The factor $-i$ included in the definitions of $\mh$ and $\me$ has been introduced to facilitate the calculation later. Using again the integral representation of the delta functions for these mean-field parameters, $\isingZ$ becomes
{\small
\begin{align}
\isingZ
&= \int \frac{d \hat{M}}{2\pi} \int \frac{d \ms d \hms}{2\pi} \int \frac{d \msm d \hmsm}{2\pi} \int \frac{d \mh d \hmh}{2\pi} \int \frac{d\me d\hme}{2\pi}
\nonumber\\
&\times\exp\left[iM\hM\!+\!i\ms\hms\!+\!i\msm\hmsm\!+\!i\mh\hmh\!+\!i\me\hme\!-\!i\hM\ms\!+\!NJ_0\ms\mh\!+\!NJ_0\msm\me\right]
\nonumber\\
&\times
\prod_i\left\{\sum_{s_i=\pm 1}\sum_{\sigma_i=\pm 1}\int \frac{d h_i d\hat{h_i}}{2\pi}\int \frac{d \eta_i d\hat{\eta_i}}{2\pi} e^{i h_i\hat{h}_i + i\eta_i\hat{\eta}_i}\right.
\nonumber\\
&\qquad\qquad\times\left.
\left[\frac{1\!-\!\sigma_i}{2}+\frac{1\!+\!\sigma_i}{2}\Theta\left(h_i\eta_i\right)\right]
e^{\frac{1}{N}\left[-i\hms s_i-i\hmsm s_i\sigma_i-i\hmh (-i\hat{h}_i)-i\hme (-i\hat{\eta}_i)\right]+\gamma\frac{1\!+\!\sigma_i}{2}}\right\}.
\end{align}
}
With the change of variables $\hM\!\to\!iN\hM$, $\hms\!\to\!iN\hms$, $\hmsm\!\to\!iN\hmsm$, $\hmh\!\to\!iN\hmh$ and $\hme\!\to\!iN\hme$, one can show that $\isingZ$ is given by
\begin{align}
\label{eq_NPsi}
\isingZ
\propto \int \frac{d \hat{M}}{2\pi} \int \frac{d \ms d \hms}{2\pi} \int \frac{d \msm d \hmsm}{2\pi} \int \frac{d \mh d \hmh}{2\pi} \int \frac{d\me d\hme}{2\pi} e^{N\isingPsi}
\end{align}
such that
\begin{align}
\isingPsi &= -M\hM-\ms\hms - \msm\hmsm - \mh\hmh - \me\hme + \hM\ms+J_0\ms\mh+J_0\msm\me
\nonumber\\
&+\log\left\{\sum_{s=\pm 1}\sum_{\sigma=\pm 1}\int \frac{d h~d\hat{h}}{2\pi}\int \frac{d \eta~d\hat{\eta}}{2\pi} e^{i h\hat{h} + i\eta\hat{\eta}}\right.
\nonumber\\
&\qquad\qquad\times\left.
\left[\frac{1\!-\!\sigma}{2}+\frac{1\!+\!\sigma}{2}\Theta\left(h\eta\right)\right]
e^{\hms s +\hmsm s\sigma+\hmh (-i\hat{h})+\hme (-i\hat{\eta})+\gamma\frac{1\!+\!\sigma}{2}}\right\},
\end{align}
where the site index $i$ is omitted as all the terms are factorized. One can then integrate $\hat{h}$ and $\hat{\eta}$ which become the delta functions $\delta(h-\hmh)$ and $\delta(\eta-\hme)$; implementing these delta functions by integrating $h$ and $\eta$, $\isingPsi$ becomes
\begin{align}
\isingPsi &= -M\hM-\ms\hms - \msm\hmsm - \mh\hmh - \me\hme + \hM\ms+J_0\ms\mh+J_0\msm\me
\nonumber\\
&+\log\left\{\sum_{s=\pm 1}\sum_{\sigma=\pm 1}
\left[\frac{1\!-\!\sigma}{2}+\frac{1\!+\!\sigma}{2}\Theta\left(\hmh\hme\right)\right]
e^{\hms s +\hmsm s\sigma+\gamma\frac{1\!+\!\sigma}{2}}\right\}.
\end{align}
We can then sum over of $s$ and $\sigma$, such that $\isingPsi$ becomes
\begin{align}
\label{eq_isingPsi}
\isingPsi =& -M\hM-\ms\hms - \msm\hmsm - \mh\hmh - \me\hme + \hM\ms+J_0\ms\mh+J_0\msm\me
\nonumber\\
&+\log\left[2\cosh(\hms-\hmsm)+2~\Theta(\hmh\hme)\cosh(\hms+\hmsm)e^\gamma\right].
\end{align}

To compute $\isingZ$, one can then make use of \req{eq_NPsi} to evaluate $\isingZ$ by the method of steepest descent, such that the integral is given by $e^{N\isingPsi}$ when $\isingPsi$ attains its maximum value. We thus differentiate $\isingPsi$ in \req{eq_isingPsi} with respect to $\hM$, the $m$ variables and the $\hat{m}$ variables (but not $M$ which is a given constant). The differentiation of \req{eq_isingPsi} with respect to $\hM$ leads to
\begin{align}
\ms = M
\end{align}
as expected and suggested by \req{eq_m}. The differentiation of \req{eq_isingPsi} with respect to the $m$ variables gives
\begin{align}
&\hms = \hM + J_0\mh
\\
&\hmsm = J_0\me
\\
&\hmh = J_0\ms
\\
&\hme = J_0\msm
\end{align}
The differentiation with respect to the $\hat{m}$ variables results in
\begin{align}
&\ms = \frac{\sinh(\hms-\hmsm)+\Theta(\hmh\hme)\sinh(\hms+\hmsm)e^\gamma}{\cosh(\hms-\hmsm)+\Theta(\hmh\hme)\cosh(\hms+\hmsm)e^\gamma},
\\
&\msm = \frac{-\sinh(\hms-\hmsm)+\Theta(\hmh\hme)\sinh(\hms+\hmsm)e^\gamma}{\cosh(\hms-\hmsm)+\Theta(\hmh\hme)\cosh(\hms+\hmsm)e^\gamma},
\\
&\mh =
\frac{\hme\delta(\hmh\hme)\sinh(\hms+\hmsm)e^\gamma}{\cosh(\hms-\hmsm)+\Theta(\hmh\hme)\cosh(\hms+\hmsm)e^\gamma},
\\
&\me =
\frac{\hmh\delta(\hmh\hme)\sinh(\hms+\hmsm)e^\gamma}{\cosh(\hms-\hmsm)+\Theta(\hmh\hme)\cosh(\hms+\hmsm)e^\gamma},
\end{align}
where $\delta(x)$ is the delta function. One can summarize all the above relations into 4 equations with 4 unknowns, namely $\hM, \msm, \mh$ and $\me$, and the following equations
\begin{align}
\label{eq_M1}
&M = \frac{\sinh(\hM+J_0\mh-J_0\me)+\Theta(M\msm)\sinh(\hM+J_0\mh+J_0\me)e^\gamma}{\cosh(\hM+J_0\mh-J_0\me)+\Theta(M\msm)\cosh(\hM+J_0\mh+J_0\me)e^\gamma},
\\
\label{eq_msm1}
&\msm = \frac{-\sinh(\hM+J_0\mh-J_0\me)+\Theta(M\msm)\sinh(\hM+J_0\mh+J_0\me)e^\gamma}{\cosh(\hM+J_0\mh-J_0\me)+\Theta(M\msm)\cosh(\hM+J_0\mh+J_0\me)e^\gamma},
\\
&\mh = \frac{\msm\delta(M\msm)\sinh(\hM+J_0\mh+J_0\me)e^\gamma}{\cosh(\hM+J_0\mh-J_0\me)+\Theta(M\msm)\cosh(\hM+J_0\mh+J_0\me)e^\gamma},
\\
&\me = \frac{M\delta(M\msm)\sinh(\hM+J_0\mh+J_0\me)e^\gamma}{\cosh(\hM+J_0\mh-J_0\me)+\Theta(M\msm)\cosh(\hM+J_0\mh+J_0\me)e^\gamma},
\label{eq_Mlast}
\end{align}
where the original argument of the step function is $J_0^2 M\msm$ and we have omitted the factor $J_0^2$ since it is always positive and does not influence the value of the step function. Although it seems difficult to solve equations (\ref{eq_M1})-(\ref{eq_Mlast}), we will show later that $\mh = \me = 0$ is a self-consistent solution. We thus put $\mh = \me = 0$ into \req{eq_M1} which leads to
\begin{align}
\label{eq_M}
&M = \tanh\hM.
\end{align}
Since $M$ is the given magnetization of the Ising model, the above equation is satisfied by
\begin{align}
\label{eq_hM}
\hM = \beta J_0 M
\end{align}
such that $M=\tanh(\beta J_0 M)$ as in the original Ising model, and the physical inverse temperature $\beta$ appears naturally even if one assumes no knowledge of the temperature $T$ in $\isingZ$ in \req{eq_isingZ}. In this case, from \req{eq_msm1} we have
\begin{align}
\label{eq_msm}
\msm = (\tanh\hM)
\left(\frac{\Theta(M\msm)e^\gamma-1}{\Theta(M\msm)e^\gamma+1}\right) = M\left(\frac{\Theta(M\msm)e^\gamma-1}{\Theta(M\msm)e^\gamma+1}\right),
\end{align}
as stated in the main paper. Finally, we use Eqs.~(S\ref{eq_hM}) and (S\ref{eq_msm}) to show that the ansatz $\mh = \me = 0$ is consistent. We note that when $M\neq 0$, i.e. in the ferromagnetic phase, $\msm\neq 0$ for all $\gamma$ except the ambiguity at $\gamma=-\ln[\Theta(0)]$ which turns out to be the singular point at cluster size $\size=0.5$ in the final solution, thus $\mh=\me\propto\delta(M\msm)=0$ in the ferromagnetic phase. For the paramagnetic phase, the magnetization $M=0$ and $\mh=\me\propto 0\cdot\delta(0)$ which do not have a well defined value. Nevertheless, putting $\mh=\me=0$ results in Eqs.~(S\ref{eq_M}) and (S\ref{eq_hM}) as obtained in the original Ising model and is thus a consistent solution in the paramagnetic phase.

We finally substitute Eqs.~(S\ref{eq_M})-(S\ref{eq_msm}) and $\mh=\me=0$ into \req{eq_isingPsi} which results in
\begin{align}
\isingPsi = -\beta J_0 M^2 + \log\left\{2\cosh(\beta J_0 M)[1+\Theta(M\msm)e^\gamma]\right\}.
\end{align}
In the limit $N\to\infty$, $\isingZ=e^{N\isingPsi}$ and is given by
\begin{align}
\label{eq_isingZ_solution}
\isingZ(\gamma, M) = A(M) [1+\Theta(M\msm)e^\gamma]^N,
\end{align}
such that
\begin{align}
A(M) = e^{-\beta N J_0 M^2}[2\cosh(\beta J_0 M)]^N
\end{align}
as stated in the main paper. By a similar calculation, one can show that $A(M)=\trace_{\{s_i\}}\delta(\sum_i s_i/N - M)$, i.e. the entropic contribution of the spin variables $\{s_i\}$.

\section{Derivation of $\skZ$}

To derive $\skZ$, we start from the replicated operator partition function
\begin{align}
\skXi(\gamma,\!\{\Ma\},\!\{\Qab\},n)
&=\trace_{\{J_{ij}\}}\trace_{\{\sia\}}\trace_{\{\sgia\}}e^{\gamma\sum_{i, \alpha}\frac{\sum_{i,\alpha}(1\!+\!\sgia)}{2}}\prod_{\alpha}w(\{\sgia\},\!\{\sia\},\!\{J_{ij}\})
\nonumber\\
&\times
\prod_{\alpha}\delta\left(\frac{\sum_i\sia}{N}\!-\!\ma\right)
\prod_{\alpha\beta}\delta\left(\frac{\sum_i\sia\sib}{N}\!-\!\qab\right) P(\mathbf{J}),
\nonumber
\end{align}
where we will use the capital letters $\Ma$ and $\Qab$ instead of $\ma$ and $\qab$ to represent the SK model order parameters to avoid confusion in subsequent derivations. Also here, we uniformly draw system configurations from those which are consistent with the order parameters that fully describe the model macroscopically. Following the expression of \req{eq_ising_step1} in the case of the Ising model, we: (i) use the integral representation of delta function to represent $\delta(\sum_i \sia/N - \Ma)$ and $\delta\left(\sum_i\sia\sib/N-\Qab\right)$; (ii) denote $\hia = \sum_{j}J_{ij}\sja$ and $\eia=\sum_{j}J_{ij}\sja \sgja$. These lead to
{\small
\begin{align}
\label{eq_sk_step1}
&\skXi
= \trace_{\{\sia\}}\trace_{\{\sgia\}}\prod_{(ij)}\left[\int dJ_{ij}\rho(J_{ij})\right]
\nonumber\\
&\quad\times
\prod_{\alpha}\left[\int \frac{d \hMa}{2\pi}e^{i \Ma\hMa - i\hMa\frac{\sum_i\sia}{N}}\right]
\prod_{\alpha\beta}\left[\int \frac{d \hQab}{2\pi}e^{i \Qab\hQab - i\hQab\frac{\sum_i\sia\sib}{N}}\right]
\nonumber\\
&\quad\times
\prod_{i\alpha}\left[\int\!\frac{d \hia d\hhia}{2\pi}\int\!\frac{d \eia d\heia}{2\pi} e^{i \hia\hhia+ i\eia\heia }\right]\prod_{(ij)}\left[e^{-i J_{ij}\sum_\alpha\left(\hhia\sja+\hhja\sia +\heia\sja \sgja+\heja\sia\sgia\right)}\right]
\nonumber\\
&\quad\times \prod_{i\alpha}\left[\frac{1-\sgia}{2}+\frac{1+\sgia}{2}\Theta\left(\hia\eia\right)\right]
e^{\gamma\frac{\sum_{i\alpha} (1+\sgia)}{2}}.
\end{align}
}
We then average the coupling disorder by integrating $J_{ij}$ over the distribution
\begin{align}
\rho(J_{ij})=\sqrt{\frac{N}{2\pi J^2}}e^{-\frac{N}{2J^2}\left(J_{ij}-\frac{J_0}{N}\right)^2}.
\nonumber
\end{align}
for each $(ij)$, to obtain
\begin{align}
&\prod_{(ij)}\left\{\int dJ_{ij}\rho(J_{ij})e^{-i J_{ij}\sum_\alpha\left(\hhia\sja+\hhja\sia +\heia\sja \sgja+\heja\sia\sgia\right)}\right\}
\nonumber\\
&\propto\prod_{(ij)}\exp\left\{-\frac{J^2}{2N}\sum_{\alpha, \beta}\left(
\hhia\sja \hhib\sjb + \hhia\sja \hhjb\sib + \hhia\sja \heib\sjb\sgjb + \hhia\sja \hejb\sib\sgib
\right.\right.
\nonumber\\
&\qquad\qquad+\hhja\sia \hhib\sjb +\hhja\sia \hhjb\sib +\hhja\sia \heib\sjb\sgjb +\hhja\sia \hejb\sib\sgib
\nonumber\\
&\qquad\qquad+\heia\sja\sgja \hhib\sjb +\heia\sja\sgja \hhjb\sib +\heia\sja\sgja \heib\sjb\sgjb +\heia\sja\sgja \hejb\sib\sgib
\nonumber\\
&\left.\qquad\qquad+\heja\sia\sgia \hhib\sjb +\heja\sia\sgia \hhjb\sib +\heja\sia\sgia \heib\sjb\sgjb +\heja\sia\sgia \hejb\sib\sgib\right)
\nonumber\\
&\left.\quad
-i\frac{J_0}{N}\sum_\alpha\left(\hhia\sja+\hhja\sia +\heia\sja \sgja+\heja\sia\sgia\right)\right\}
\nonumber\\
&=\exp\left\{\frac{NJ^2}{4}\sum_{\alpha, \beta}\left[2\qssab\qhhab + 2\left(\qshab\right)^2 + 4\qheab\qssmab + 4\qhsmab\qseab + 2\qeeab\qssmmab + 2\left(\qesmab\right)^2 \right]
\right.
\nonumber\\
&\left.
\quad+\frac{NJ_0}{2}\sum_\alpha \left[2\msa\mha + 2\msma\mea\right]
\right\}
\end{align}
where we have neglected terms of $O(N)$ in the last line and keep only terms of $O(N^2)$. We then define mean-field parameters, as in Eqs.~(S\ref{eq_m})-(S\ref{eq_me}) in the case of the Ising model, to be
\begin{align}
\label{eq_ma}
&\msa = \frac{1}{N}\sum_i \sia,
\\
&\msma = \frac{1}{N}\sum_i \sia \sgia,
\\
&\mha = -\frac{i}{N}\sum_i \hhia,
\\
\label{eq_mea}
&\mea = -\frac{i}{N}\sum_i \heia.
\end{align}
We also define a set of mean-field parameters to account for the correlation between replica,
\begin{align}
\label{eq_qssab}
&\qssab = \frac{1}{N}\sum_i \sia\sib,
\\
&\qshab = \frac{i}{N}\sum_i \sia\hhib,
\\
&\qseab = \frac{i}{N}\sum_i \sia\heib,
\\
\label{eq_qssmab}
&\qssmab = \frac{1}{N}\sum_i \sia\sib\sgib,
\\
&\qhhab = -\frac{1}{N}\sum_i \hhia\hhib,
\\
&\qheab = -\frac{1}{N}\sum_i \hhia\heib,
\\
&\qhsmab = \frac{i}{N}\sum_i \hhia\sgib,
\\
&\qeeab = -\frac{1}{N}\sum_i \heia\heib,
\\
&\qesmab = \frac{i}{N}\sum_i \heia\sgib,
\\
&\qesmab = \frac{1}{N}\sum_i \sia\sib\sgia\sgib~.
\end{align}
We note that $\msa$ and $\qssab$ are indeed equivalent to $\Ma$ and $\Qab$, respectively, and hence we expect to obtain $\msa=\Ma$ and $\qssab=\Qab$ in the derivation. The above mean-field parameters are introduced in $\skXi$ by the integral representation of delta functions, which lead to
{\small
\begin{align}
\skXi
=&\prod_{\alpha}\left[\int \frac{d \hMa}{2\pi}
\int \frac{d \msa d \hmsa}{2\pi} \int \frac{d \msma d \hmsma}{2\pi} \int \frac{d \mha d \hmha}{2\pi} \int \frac{d\mea d\hmea}{2\pi}
\right]
\nonumber\\
&\times\prod_{\alpha\beta}\left[\int \frac{d \hQab}{2\pi}
\int \frac{d \qssab d \hqssab}{2\pi} \int \frac{d \qshab d \hqshab}{2\pi} \int \frac{d \qseab d \hqseab}{2\pi} \int \frac{d \qssmab d \hqssmab}{2\pi} \int \frac{d \qhhab d \hqhhab}{2\pi}
\right.
\nonumber\\
&\left.
\qquad\times\int \frac{d \qshab d \hqshab}{2\pi}
\int \frac{d \qhsmab d \hqhsmab}{2\pi} \int \frac{d \qeeab d \hqeeab}{2\pi} \int \frac{d \qesmab d \hqesmab}{2\pi} \int \frac{d \qssmmab d \hqssmmab}{2\pi}
\right]
\nonumber\\
&\times\exp\left[i\sum_\alpha\left( \Ma\hMa  - \hMa\msa  + \msa\hmsa + \msma\hmsma + \mha\hmha + \mea\hmea\right)
\right.
\nonumber\\
&
\qquad+ NJ_0\sum_\alpha\left(\msa\mha + \msma\mea\right)+i\sum_{\alpha, \beta}\left(\Qab\hQab-\hQab\qssab +\qssab\hqssab + \qshab\hqshab
\right.
\nonumber\\
&\left.
\qquad+ \qseab\hqseab +\qssmab\hqssmab + \qhhab\hqhhab + \qheab\hqheab + \qhsmab\hqhsmab +\qeeab\hqeeab +\qesmab\hqesmab + \qssmmab\hqssmmab\right)
\nonumber\\
&\left.
N\frac{J^2}{2}\sum_{\alpha, \beta}\left(\qssab\qhhab + \left(\qshab\right)^2 + 2\qheab\qssmab + 2\qhsmab\qseab + \qeeab\qssmmab + \left(\qesmab\right)^2
\right)\right]
\nonumber\\
&\times
\prod_i\left\{\prod_\alpha\left[\sum_{\sia=\pm 1}\sum_{\sgia=\pm 1}\int \frac{d \hia d\hhia}{2\pi}\int \frac{d \eia d\heia}{2\pi} e^{i \hia\hhia + i\eia\heia}
\right.\right.
\nonumber\\
&\qquad\times
\left.\left(\frac{1\!-\!\sgia}{2}+\frac{1\!+\!\sgia}{2}\Theta\left(\hia\eia\right)\right)\right]
\nonumber\\
&\qquad\times
\exp\left[-\frac{i}{N}\sum_\alpha\left(\hmsa \sia+\hmsma \sia\sgia+\hmha (-i\hhia)+\hmea (-i\heia)\right)
\right.
\nonumber\\
&\qquad\qquad
-\frac{i}{N}\sum_{\alpha. \beta}\Big(\hqssab\sa\sbb
+\hqshab\sa(i\hhb)+\hqseab\sia(i\hhb) +\hqssmab\sa\sbb\sgb
\nonumber\\
&\qquad\qquad
+ \hqhhab(i\hhia)(i\hhib)
+ \hqheab(i\hia)(i\eib) + \hqhsmab(i\hhia)\sib\sgib + \hqeeab(i\heia)(i\heib)
\nonumber\\
&\qquad\qquad
\left.\left.
+ \hqesmab(i\heia)\sib\sgib
+ \hqssmmab\sia\sib\sgia\sgib
\Big)+\gamma\sum_\alpha\frac{1\!+\!\sgia}{2}\right]\right\}.
\end{align}
}

We proceed by: (i) making the change of variables $\hMa\to\!iN\hMa$, $\hQab\to\!iN\hQab$, and similar changes of variables for all the other variables of $\hqab$ and $\hma$; (ii) assuming replica symmetry (RS) such that for all $\alpha$, $\Ma=M$, $\hMa=\hM$, and similarly for other $\ma$ and $\hma$ variables, i.e. $\ma=m$ and $\hma=\hm$; (iii) for $\alpha\neq\beta$, $\Qab=Q$ and $\hQab=\hQ$, and similarly for other $\qab$ and $\hqab$, i.e. $\qab=q$ and $\hqab=\hq$; (iv) for $\alpha=\beta$, $\Qab=1$, $\hQab=\hG$, and for other variables of $\qab$ and $\hqab$ we assume $\qab=\qzero$ and $\hqab=\hg$. In this case, one can show that
\begin{align}
\skXi \propto e^{Nn\skPsi}
\end{align}
such that
{\small
\begin{align}
\label{eq_skPsi}
n\skPsi =&
-n\Big( M\hM - \hM\ms  + \ms\hms + \msm\hmsm + \mh\hmh + \me\hme\Big)
+ nJ_0\Big(\ms\mh + \msm\me\Big)
\nonumber\\
&
-n(n-1)
\Big(Q\hQ-\hQ\qss +\qss\hqss + \qsh\hqsh + \qse\hqse +\qssm\hqssm + \qhh\hqhh + \qhe\hqhe
\nonumber\\
&
+ \qhsm\hqhsm +\qee\hqee +\qesm\hqesm + \qssmm\hqssmm\Big) - n\Big(\hG-\hG\gss +\gss\hgss + \gsh\hgsh + \gse\hgse
\nonumber\\
&
+\gssm\hgssm  + \ghh\hghh + \ghe\hghe + \ghsm\hghsm +\gee\hgee +\gesm\hgesm + \gssmm\hgssmm\Big)
\nonumber\\
&+\frac{n(n-1)J^2}{2}\Big(\qss\qhh + \left(\qsh\right)^2 + 2\qhe\qssm + 2\qhsm\qse + \qee\qssmm + \left(\qesm\right)^2 \Big)
\nonumber\\
&+\frac{nJ^2}{2}\Big(\gss\ghh + \left(\gsh\right)^2 + 2\ghe\gssm + 2\ghsm\gse + \gee\gssmm + \left(\gesm\right)^2 \Big)+n\Phi
\end{align}
}
and
{\small
\begin{align}
n\Phi =& \log\left\{\prod_\alpha\left[\sum_{\sa=\pm 1}\sum_{\sga=\pm 1}\int \frac{d \ha d\hha}{2\pi}\int \frac{d \ea d\hea}{2\pi} e^{i \ha\hha + i\ea\hea}
\left(\frac{1\!-\!\sga}{2}+\frac{1\!+\!\sga}{2}\Theta\left(\ha\ea\right)\right)\right]
\right.
\nonumber\\
&\qquad\times
\exp\left[\left(\hms \sum_\alpha\sa+\hmsm \sum_\alpha\sa\sga+\hmh\sum_\alpha (-i\hha)+\hme\sum_\alpha (-i\hea)\right)
\right.
\nonumber\\
&\qquad
+\left(\hqss\Big(\suma\sa\Big)^2+\hqsh\suma\sa\sumb i\hhb
+\hqse\suma\sa\sumb i\heb +\hqssm\suma\sa\sumb\sbb\sgb
\right.
\nonumber\\
&\qquad\qquad
+ \hqhh\Big(\suma i\hha\Big)^2 + \hqhe\suma i\hha\sumb i\heb
+ \hqhsm\suma i\hha\sumb\sbb\sgb
\nonumber\\
&\qquad\qquad
\left.
 + \hqee\Big(\suma i\hea\Big)^2 + \hqesm \suma i\hea\sumb\sbb\sgb + \hqssmm\Big(\suma\sa\sga\Big)^2
\right)
\nonumber\\
&\qquad
+\left(n(\hgss\!-\!\hqss)+(\hgsh\!-\!\hqsh)\suma\sa (i\hha)
+(\hgse\!-\!\hqse)\suma\sa (i\hea) +(\hgssm\!-\!\hqssm)\suma\sga
\right.
\nonumber\\
&\qquad\qquad
+ (\hghh\!-\!\hqhh)\suma (i\hha)^2 + (\hghe\!-\!\hqhe)\suma (i\hha) (i\hea)
+ (\hghsm-\hqhsm)\suma (i\hha)\sa\sga
\nonumber\\
&\qquad\qquad
\left.
 + (\hgee\!-\!\hqee)\suma (i\hea)^2 + (\hgesm\!-\!\hqesm) \suma (i\hea)\sa\sga + n(\hgssmm\!-\!\hqssmm)
\right)
\nonumber\\
&\qquad
\left.\left.
+\gamma\sum_\alpha\frac{1\!+\!\sgia}{2}\right]\right\}.
\label{eq:Phi}
\end{align}
}
One can see that all the terms in \req{eq:Phi} are factorized with respect to the replica index $\alpha$ except the terms from the 3rd to the 5th line. In addition, we also have to linearize $(i\hha)^2$ and $(i\hea)^2$, and decouple $(i\hha) (i\hea)$, which will finally become the delta functions of $\ha$ and $\ea$ by integrating the corresponding $\hha$ and $\hea$ respectively. To achieve this we re-write $n\Phi$ as
\begin{align}
n\Phi &= n(\hgss\!-\!\hqss) + n(\hgssmm\!-\!\hqssmm)
\nonumber\\
&
+\log\left\{\prod_\alpha\left[\sum_{\sa=\pm 1}\sum_{\sga=\pm 1}\int \frac{d \ha d\hha}{2\pi}\int \frac{d \ea d\hea}{2\pi} e^{i \ha\hha + i\ea\hea}
\left(\frac{1\!-\!\sga}{2}+\frac{1\!+\!\sga}{2}\Theta\left(\ha\ea\right)\right)\right]
\right.
\nonumber\\
&\qquad\times
\exp\left[\hms \sum_\alpha\sa+\hmsm \sum_\alpha\sa\sga+\hmh\sum_\alpha (-i\hha)+\hme\sum_\alpha (-i\hea)
\right.
\nonumber\\
&\qquad
+\frac{1}{2}\vec{w}^T\cdot\matQ\cdot\vec{w}
+ \frac{1}{2}\suma \vec{y}^T_\alpha\cdot\matG\cdot\vec{y}_\alpha +(\hgsh\!-\!\hqsh)\suma\sa (i\hha)
+(\hgse\!-\!\hqse)\suma\sa (i\hea)
\nonumber\\
&\qquad
+(\hgssm\!-\!\hqssm)\suma\sga + (\hghsm-\hqhsm)\suma (i\hha)\sa\sga
+ (\hgesm\!-\!\hqesm) \suma (i\hea)\sa\sga
\nonumber\\
&\qquad
\left.\left.
+\gamma\sum_\alpha\frac{1\!+\!\sga}{2}\right]\right\}.
\end{align}
where
\begin{align}
\matQ =
\begin{pmatrix}
2\hqss & \hqsh & \hqse & \hqssm
\\
\hqsh & 2\hqhh & \hqhe & \hqhsm
\\
\hqse & \hqhe & 2\hqee & \hqesm
\\
\hqssm & \hqhsm & \hqesm & 2\hqssmm
\end{pmatrix},
\qquad\qquad
\vec{w} =
\begin{pmatrix}
\suma \sa
\\
\suma i\hha
\\
\suma i\hea
\\
\suma \sa\sga
\end{pmatrix},
\end{align}
and
\begin{align}
\matG =
\begin{pmatrix}
2(\hghh-\hqhh) & \hghe-\hqhe
\\
\hghe-\hqhe & 2(\hgee - \hqee)
\end{pmatrix},
\qquad\qquad
\vec{y}_\alpha =
\begin{pmatrix}
i\hha
\\
i\hea
\end{pmatrix}.
\end{align}
We can then adopt multivariate Gaussian integrals to linearize $\vec{w}^T\cdot\matQ\cdot\vec{w}$ and $\vec{y}^T_\alpha\cdot\matG\cdot\vec{y}_\alpha$, such that $n\Phi$ becomes
\begin{align}
n\Phi &= n(\hgss\!-\!\hqss) + n(\hgssmm\!-\!\hqssmm) + \log\left\{\frac{1}{4\pi^2\sqrt{|\matQ|}}\int \dz e^{-\frac{1}{2}\vz^T\matQ^{-1}\vz}
\right.
\nonumber\\
&\times\exp n\log\left[\frac{1}{2\pi\sqrt{|\matG|}}\int \dx e^{-\frac{1}{2}\vx^T\matG^{-1}\vx}\sum_{s=\pm 1}\sum_{\sigma=\pm 1}\int \frac{d h d\hh}{2\pi}\int \frac{d \eta d\he}{2\pi} e^{i h\hh + i\eta\he}
\right.
\nonumber\\
&\qquad\times
\left(\frac{1\!-\!\sigma}{2}+\frac{1\!+\!\sigma}{2}\Theta\left(h\eta\right)\right)
\exp\bigg(z_1s + z_2(i\hh) + z_3(i\he) + z_4(s\sigma) + x_1(i\hh) + x_2(i\he)
\nonumber\\
&\qquad
+ \hms s+\hmsm s\sigma +\hmh (-i\hh)+\hme (-i\he)
+(\hgsh\!-\!\hqsh)s (i\hh)
+(\hgse\!-\!\hqse)s (i\he)
\nonumber\\
&\qquad
+(\hgssm\!-\!\hqssm)\sigma + (\hghsm-\hqhsm) (i\hh)s\sigma
+ (\hgesm\!-\!\hqesm) (i\he)s\sigma
+\gamma\frac{1\!+\!\sigma}{2}
\bigg)
\Bigg]\Bigg\}.
\end{align}
We continue the calculation by: (i) expanding the exponential function in the 2nd line as a power of $n$, such that in the limit of small $n$ one can make use of  $\log[1+nC+O(n^2)+...]\approx nC$ to simplify the expression; (ii) collecting terms with factors of $i\hh$ and $i\he$, and the integration of $\hh$ and $\he$ gives rise to the delta functions
\begin{align}
\delta\left[h+\left(z_2+x_1+(\hgsh\!-\!\hqsh)s +  (\hghsm-\hqhsm) s\sigma - \hmh \right)\right],
\\
\delta\left[\eta+\left(z_3+x_2+(\hgse\!-\!\hqse)s +  (\hgesm-\hqesm) s\sigma - \hme \right)\right].
\end{align}
Integrating $h$ and $\eta$ lead to
\begin{align}
n\Phi &= n(\hgss\!-\!\hqss) + n(\hgssmm\!-\!\hqssmm) + \frac{n}{4\pi^2\sqrt{|\matQ|}}\int \dz e^{-\frac{1}{2}\vz^T\matQ^{-1}\vz}
\nonumber\\
&\times
\log\Bigg\{\frac{1}{2\pi\sqrt{|\matG|}}\int \dx e^{-\frac{1}{2}\vx^T\matG^{-1}\vx}\sum_{s=\pm 1}\sum_{\sigma=\pm 1}
\nonumber\\
&\times
\bigg(\frac{1\!-\!\sigma}{2}+\frac{1\!+\!\sigma}{2}\Theta\Big[\left(z_2\!+\!x_1+\!(\hgsh\!-\!\hqsh)s\!+\!(\hghsm-\hqhsm) s\sigma\!-\!\hmh\right)
\nonumber\\
&
\qquad\qquad\qquad\qquad\qquad\qquad
\times
\left(z_3\!+\!x_2\!+\!(\hgse\!-\!\hqse)s\!+\!(\hgesm-\hqesm) s\sigma\!-\!\hme\right)\Big]\bigg)
\nonumber\\
&
\qquad\qquad\qquad
\times
\exp\left(z_1s + z_4 s\sigma + \hms s + \hmsm s\sigma +(\hgssm\!-\!\hqssm)\sigma +\gamma\frac{1\!+\!\sigma}{2}\right)
\Bigg\}
\end{align}
Finally, we sum over $s$ and $\sigma$ to give
{\small
\begin{align}
\label{eq_Phi}
\Phi &= (\hgss\!-\!\hqss) + (\hgssmm\!-\!\hqssmm) + \frac{1}{4\pi^2\sqrt{|\matQ|}}\int \dz e^{-\frac{1}{2}\vz^T\matQ^{-1}\vz}
\nonumber\\
&\times\log\Bigg\{2\cosh[z_1\!-\!z_4\!+\!\hms\!-\!\hmsm]e^{-(\hgssm\!-\!\hqssm)}
\nonumber\\
&\quad+e^{-z_1-z_4-\hms-\hmsm+(\hgssm\!-\!\hqssm)+\gamma}
\frac{1}{2\pi\sqrt{|\matG|}}\int \dx e^{-\frac{1}{2}\vx^T\matG^{-1}\vx}
\nonumber\\
&\qquad\times\Theta\Big[\Big(z_2\!+\!x_1-\!(\hgsh\!-\!\hqsh)\!-\!(\hghsm-\hqhsm) \!-\!\hmh\Big)\Big(z_3\!+\!x_2\!-\!(\hgse\!-\!\hqse)\!-\!(\hgesm-\hqesm) \!-\!\hme\Big)\Big]
\nonumber\\
&\quad+ e^{z_1+z_4+\hms+\hmsm+(\hgssm\!-\!\hqssm)+\gamma}
\frac{1}{2\pi\sqrt{|\matG|}}\int \dx e^{-\frac{1}{2}\vx^T\matG^{-1}\vx}
\nonumber\\
&\qquad\times\Theta\Big[\Big(z_2\!+\!x_1+\!(\hgsh\!-\!\hqsh)\!+\!(\hghsm-\hqhsm) \!-\!\hmh\Big)\Big(z_3\!+\!x_2\!+\!(\hgse\!-\!\hqse)\!+\!(\hgesm-\hqesm) \!-\!\hme\Big)\Big]
\Bigg\}
\end{align}
}
By using the above expression of $\Phi$ and \req{eq_skPsi}, we can write $\lim_{n\to 0}\skPsi$ as
{\small
\begin{align}
\label{eq_skPsi2}
\lim_{n\to 0}\skPsi
&=
-\Big( M\hM - \hM\ms  + \ms\hms + \msm\hmsm + \mh\hmh + \me\hme\Big)
+ J_0\Big(\ms\mh + \msm\me\Big)
\nonumber\\
&
+\Big(Q\hQ-\hQ\qss +\qss\hqss + \qsh\hqsh + \qse\hqse +\qssm\hqssm + \qhh\hqhh + \qhe\hqhe
\nonumber\\
&
+ \qhsm\hqhsm +\qee\hqee +\qesm\hqesm + \qssmm\hqssmm\Big) - \Big(\hG-\hG\gss +\gss\hgss + \gsh\hgsh + \gse\hgse
\nonumber\\
&
+\gssm\hgssm  + \ghh\hghh + \ghe\hghe + \ghsm\hghsm +\gee\hgee +\gesm\hgesm + \gssmm\hgssmm\Big)
\nonumber\\
&-\frac{J^2}{2}\Big(\qss\qhh + \left(\qsh\right)^2 + 2\qhe\qssm + 2\qhsm\qse + \qee\qssmm + \left(\qesm\right)^2 \Big)
\nonumber\\
&+\frac{J^2}{2}\Big(\gss\ghh + \left(\gsh\right)^2 + 2\ghe\gssm + 2\ghsm\gse + \gee\gssmm + \left(\gesm\right)^2 \Big)+ (\hgss-\hqss)
\nonumber\\
&+(\hgssmm-\hqssmm) + \int\Dz\log
\Bigg\{2\cosh[z_1\!-\!z_4\!+\!\hms\!-\!\hmsm]e^{-(\hgssm\!-\!\hqssm)} + e^{(\hgssm\!-\!\hqssm)\!+\!\gamma}
\nonumber\\
&\times\bigg[e^{-z_1-z_4-\hms-\hmsm}
\Omega_{-}(z_2, z_3, \hmh, \hme, \mathbf{\hq}, \mathbf{\hg})
+ e^{z_1+z_4+\hms+\hmsm}
\Omega_{+}(z_2, z_3, \hmh, \hme, \mathbf{\hq}, \mathbf{\hg})
\bigg]\Bigg\},
\end{align}
}
where $\int \Dz$ represents the multivariate Gaussian integration
\begin{align}
\frac{1}{4\pi^2\sqrt{|\matQ|}}\int \dz e^{-\frac{1}{2}\vz^T\matQ^{-1}\vz},
\end{align}
and the functions $\Omega_{\pm}$ are given by
\begin{align}
\label{eq_omega}
&\Omega_\pm(z_2, z_3, \hmh, \hme, \mathbf{\hq}, \mathbf{\hg}) = \frac{1}{2\pi\sqrt{|\matG|}}\int \dx e^{-\frac{1}{2}\vx^T\matG^{-1}\vx}
\nonumber\\
&\quad\times\Theta\Big[\Big(z_2\!+\!x_1\pm\!(\hgsh\!-\!\hqsh)\!\pm\!(\hghsm-\hqhsm) \!-\!\hmh\Big)\Big(z_3\!+\!x_2\!\pm\!(\hgse\!-\!\hqse)\!\pm\!(\hgesm-\hqesm) \!-\!\hme\Big)\Big].
\end{align}
such that $\mathbf{\hq}$ and $\mathbf{\hg}$ represent vectors of the variables of $\hq$ and $\hg$ respectively.

\subsection{Saddle point equations}

Since $\skXi\propto e^{Nn\skPsi}$, one can evaluate
\begin{align}
\frac{1}{N}\ln \skZ=\frac{1}{N}\lim_{n\to0}\frac{\skXi-1}{n}=\lim_{n\to 0}\skPsi^*
\end{align}
by the method of steepest descent such that $\skPsi^*$ corresponds to the extremum of $\skPsi$ with respect to $\hM, \hQ$ and all the 48 variables of $m$, $\hm$, $q$, $\hq$, $\qzero$ and $\hat{\qzero}$. Indeed, all the conjugate variable $\hm$, $\hq$ and $\hat{\qzero}$ can be extremized and expressed in terms of the variables of $m$, $q$ and $\qzero$. To facilitate the presentation of the saddle point equations, we will denote the function in the logarithm of \req{eq_skPsi2} by $\cF(\vz, \mathbf{\hm}, \mathbf{\hq}, \mathbf{\hg})$, i.e.
\begin{align}
&\cF(\vz, \mathbf{\hm}, \mathbf{\hq}, \mathbf{\hg}) = \Bigg\{2\cosh[z_1\!-\!z_4\!+\!\hms\!-\!\hmsm]e^{-(\hgssm\!-\!\hqssm)} + e^{(\hgssm\!-\!\hqssm)+\gamma}
\nonumber\\
&\quad\times\bigg[e^{-z_1-z_4-\hms-\hmsm}
\Omega_{-}(z_2, z_3, \hmh, \hme, \mathbf{\hq}, \mathbf{\hg})
+ e^{z_1+z_4+\hms+\hmsm}
\Omega_{+}(z_2, z_3, \hmh, \hme, \mathbf{\hq}, \mathbf{\hg})
\bigg]\Bigg\}
\end{align}

We first differentiate the above expression with respect to $\hM$, $\hQ$ and $\hG$ and obtain the expected relations:
\begin{align}
&\ms = M,
\\
&\qss = Q,
\\
&\gss = 1.
\end{align}
We then differentiate \req{eq_skPsi2} with respect to $\ms, \msm, \mh$ and $\me$ to obtain the following relations which are identical to Eqs.~(\ref{eq_m})-(\ref{eq_me}) of the Ising model
\begin{align}
&\hms = \hM + J_0\mh,
\\
&\hmsm = J_0\me,
\\
&\hmh = J_0\ms,
\\
&\hme = J_0\msm.
\end{align}
Differentiating with respect to the individual variables of $q$ and $c$ gives
\begin{align}
&\hqss = \hQ + \frac{J^2}{2}\qhh, &&\hgss = \hG + \frac{J^2}{2}\ghh
\\
&\hqsh = J^2\qsh, &&\hgsh = J^2\gsh
\\
&\hqse = J^2\qhsm, &&\hgse = J^2\ghsm
\\
&\hqssm = J^2\qhe, &&\hgssm = J^2\ghe
\\
&\hqhh = \frac{J^2}{2}\qss, &&\hghh = \frac{J^2}{2}\gss
\\
&\hqhe = J^2\qssm, &&\hghe = J^2\gssm
\\
&\hqhsm = J^2\qse, &&\hghsm = J^2\gse
\\
&\hqee = \frac{J^2}{2}\qssmm, &&\hgee = \frac{J^2}{2}\gssmm
\\
&\hqesm = J^2\qesm, &&\hgesm = J^2\gesm
\\
&\hqssmm = \frac{J^2}{2}\qee, &&\hgssmm = \frac{J^2}{2}\gee
\end{align}

The remaining tasks are to differentiate \req{eq_skPsi2} with respect to individual variables of $\hm$, $\hq$ and $\hg$, which involve differentiating the complicated function $\Phi$ in \req{eq_Phi}. We first differentiate $\hms$ and $\hmsm$ which give us
\begin{align}
\label{eq_skms}
\ms &= \int\Dz\frac{1}{\cF(\vz, \mathbf{\hm}, \mathbf{\hq}, \mathbf{\hg})}\Bigg\{2\sinh[z_1\!-\!z_4\!+\!\hms\!-\!\hmsm]e^{-(\hgssm\!-\!\hqssm)} + e^{(\hgssm\!-\!\hqssm)+\gamma}
\nonumber\\
&\quad\times\bigg[-e^{-z_1-z_4-\hms-\hmsm}
\Omega_{-}(z_2, z_3, \hmh, \hme, \mathbf{\hq}, \mathbf{\hg})
+ e^{z_1+z_4+\hms+\hmsm}
\Omega_{+}(z_2, z_3, \hmh, \hme, \mathbf{\hq}, \mathbf{\hg})
\bigg]
\Bigg\}
\\
\msm &= \int\Dz\frac{1}{\cF(\vz, \mathbf{\hm}, \mathbf{\hq}, \mathbf{\hg})}\Bigg\{-2\sinh[z_1\!-\!z_4\!+\!\hms\!-\!\hmsm]e^{-(\hgssm\!-\!\hqssm)} + e^{(\hgssm\!-\!\hqssm)+\gamma}
\nonumber\\
&\quad\times\bigg[-e^{-z_1-z_4-\hms-\hmsm}
\Omega_{-}(z_2, z_3, \hmh, \hme, \mathbf{\hq}, \mathbf{\hg})
+ e^{z_1+z_4+\hms+\hmsm}
\Omega_{+}(z_2, z_3, \hmh, \hme, \mathbf{\hq}, \mathbf{\hg})
\bigg]
\Bigg\}.
\end{align}
We remark that $\ms=M$ is known in \req{eq_skms} and one should instead extract the value of $\hms$ from \req{eq_skms}. The differentiation of \req{eq_skPsi2} with respect to $\hmh$ and $\hme$ involves differentiating the step functions in $\Omega_{\pm}$ in \req{eq_omega} and should be taken with extra care. We first differentiate $\Omega_{\pm}$ with respect to $\hmh$
\begin{align}
\frac{\partial \Omega_\pm}{\partial \hmh} &= \frac{-1}{2\pi\sqrt{|\matG|}}\int \dx e^{-\frac{1}{2}\vx^T\matG^{-1}\vx}\Big(z_3\!+\!x_2\!\pm\!(\hgse\!-\!\hqse)\!\pm\!(\hgesm-\hqesm) \!-\!\hme\Big)
\nonumber\\
&\qquad\times\delta\Big[\Big(z_2\!+\!x_1\pm\!(\hgsh\!-\!\hqsh)\!\pm\!(\hghsm-\hqhsm) \!-\!\hmh\Big)\Big(z_3\!+\!x_2\!\pm\!(\hgse\!-\!\hqse)\!\pm\!(\hgesm-\hqesm) \!-\!\hme\Big)\Big].
\nonumber\\
\qquad
&=\left.
\frac{-1}{2\pi\sqrt{|\matG|}}\int d x_2 e^{-\frac{1}{2}\vx^T\matG^{-1}\vx}\sgn\Big(z_3\!+\!x_2\!\pm\!(\hgse\!-\!\hqse)\!\pm\!(\hgesm-\hqesm) \!-\!\hme\Big)\right|_{x_1=x_1^{\pm}}
\end{align}
where we arrive at the last line by integrating $x_1$ in the delta function, such that $x_1$ in the final expression is substituted by  $x_1^{\pm}=-z_2\mp\!(\hgsh\!-\!\hqsh)\!\mp\!(\hghsm-\hqhsm)\!+\!\hmh$. One can further simplify the above expression using the properties of multivariate Gaussian distribution and the definition of error function. If we denote the element in the $i$-th row and $j$-th of $\matG$ by $v_{ij}$, the above expression becomes
\begin{align}
\frac{\partial \Omega_\pm}{\partial \hmh} &= \frac{-e^{-\frac{\left(x_1^{\pm}\right)^2}{2v_{11}}}}{2\pi\sqrt{v_{11}\left(v_{22}-\frac{v_{12}^2}{v_{11}}\right)}}
\int d x_2 e^{-\frac{\left(x_2-\frac{v_{12}x_1^{\pm}}{v_{11}}\right)^2}{2\left(v_{22}-\frac{v_{12}^2}{v_{11}}\right)}}
\sgn\Big(z_3\!+\!x_2\!\pm\!(\hgse\!-\!\hqse)\!\pm\!(\hgesm-\hqesm)\!-\!\hme\Big)
\nonumber\\
&=\frac{-e^{-\frac{\left(x_1^{\pm}\right)^2}{2v_{11}}}}{2\pi\sqrt{v_{11}}}
{\rm erf}\left(\frac{z_3\pm(\hgse\!-\!\hqse)\pm(\hgesm-\hqesm)-\hme+\frac{v_{12}x_1^{\pm}}{v_{11}}}{\sqrt{2\left(v_{22}-\frac{v_{12}^2}{v_{11}}\right)}}\right).
\end{align}
where ${\rm erf}(x)$ is the standard error function. Similarly, the differentiation of $\Omega_{\pm}$ by $\hme$ is given by
\begin{align}
\frac{\partial \Omega_\pm}{\partial \hme} &= \frac{-e^{-\frac{\left(x_2^{\pm}\right)^2}{2v_{22}}}}{2\pi\sqrt{v_{22}}}
{\rm erf}\left(\frac{z_2\pm(\hgsh\!-\!\hqsh)\pm(\hghsm-\hqhsm)-\hmh+\frac{v_{12}x_2^{\pm}}{v_{22}}}{\sqrt{2\left(v_{11}-\frac{v_{12}^2}{v_{22}}\right)}}\right).
\end{align}
where $x_2^{\pm}=-z_3\mp\!(\hgse\!-\!\hqse)\!\mp\!(\hgesm-\hqesm)\!+\!\hme$. Finally, we can differentiate \req{eq_skPsi2} with respect to $\hmh$ and $\hme$ to obtain an expression for $\mh$ and $\me$ in terms of $\partial \Omega_{\pm}/\partial\hmh$ and $\partial \Omega_{\pm}/\partial\hme$
\begin{align}
\mh &= \int\Dz\frac{1}{\cF(\vz, \mathbf{\hm}, \mathbf{\hq}, \mathbf{\hg})}\Bigg\{e^{(\hgssm\!-\!\hqssm)+\gamma}
\bigg[-e^{-z_1-z_4-\hms-\hmsm}
\frac{\partial\Omega_{-}}{\partial\hmh}
- e^{z_1+z_4+\hms+\hmsm}
\frac{\partial\Omega_{+}}{\partial\hmh}
\bigg]
\Bigg\}
\\
\me &= \int\Dz\frac{1}{\cF(\vz, \mathbf{\hm}, \mathbf{\hq}, \mathbf{\hg})}\Bigg\{e^{(\hgssm\!-\!\hqssm)+\gamma}
\bigg[-e^{-z_1-z_4-\hms-\hmsm}
\frac{\partial\Omega_{-}}{\partial\hme}
- e^{z_1+z_4+\hms+\hmsm}
\frac{\partial\Omega_{+}}{\partial\hme}
\bigg]
\Bigg\}
\end{align}

We continue to differentiate with respect to the $\hq$ variables. One can make use of the following lemma to simplify the calculations.

\vspace{0.5cm}
\noindent
\textbf{\emph{Lemma}} Given a symmetric $m\times m$ matrix $\matQ$, if we denote $u_{ij}$ to be the element of $\matQ$ in the $i$-th row and $j$-th column, then
\begin{align}
&\frac{\partial}{\partial u_{ij}}\int\Dz f(\vz) = \frac{1}{(2\pi)^m}\int d z^m e^{-\frac{1}{2}\vz^T\matQ^{-1}\vz} f(\vz)
\left(\frac{\partial}{\partial u_{ij}}\frac{1}{\sqrt{|\matQ|}}
-\frac{1}{2\sqrt{|\matQ|}}\frac{\partial}{\partial u_{ij}}\vz^T\matQ^{-1}\vz\right)
\nonumber\\
&=
\begin{cases}
\displaystyle
\frac{1}{2}\int\Dz f(\vz)\left\{-(\matQ^{-1})_{ii}+\Big[z_1(\matQ^{-1})_{1i}+z_2(\matQ^{-1})_{2i}+\dots+z_m(\matQ^{-1})_{mi}\Big]^2\right\}, & i=j
\\
\\
\displaystyle
\int\Dz f(\vz)\Bigg\{-(\matQ^{-1})_{ij}+\Big[z_1(\matQ^{-1})_{1i}+z_2(\matQ^{-1})_{2i}+\dots+z_m(\matQ^{-1})_{mi}\Big]
\\
\displaystyle
\qquad\qquad\qquad\qquad\qquad\qquad
\times\Big[z_1(\matQ^{-1})_{1j}+z_2(\matQ^{-1})_{2j}+\dots+z_m(\matQ^{-1})_{mj}\Big]
\Bigg\}, & i\neq j
\end{cases}
\end{align}
where we have made use of the relation
\begin{align}
\frac{\partial \matQ^{-1}}{\partial u_{ij}} = -\matQ^{-1} \frac{\partial \matQ}{\partial u_{ij}}\matQ^{-1}
\end{align}
for the second term in the curly brackets. For $f(\vz)$ to be a constant, one can show that $\frac{\partial}{\partial u_{ij}}\int\Dz=0$ for all $i$ and $j$ by the above lemma.
\vspace{0.5cm}

To continue the calculation, we make use of the above lemma and denote
{\small
\begin{align}
&D_{ij}=& \nonumber \\ &\phantom{=}&\!\!\!\!\!\!\!\!\!\!\!\!
\begin{cases}
\displaystyle
\frac{1}{2}\int\Dz \log\cF(\vz, \mathbf{\hm}, \mathbf{\hq}, \mathbf{\hg})
\left\{-(\matQ^{-1})_{ii}+\Big[z_1(\matQ^{-1})_{i1}+z_2(\matQ^{-1})_{i2}+z_3(\matQ^{-1})_{i3}+z_4(\matQ^{-1})_{i4}\Big]^2\right\},
\\
\hspace{14cm}
i=j
\\
\\
\displaystyle
\int\Dz \log\cF(\vz, \mathbf{\hm}, \mathbf{\hq}, \mathbf{\hg})
\Bigg\{-(\matQ^{-1})_{ij}+\Big[z_1(\matQ^{-1})_{i1}+z_2(\matQ^{-1})_{i2}+z_3(\matQ^{-1})_{i3}+z_4(\matQ^{-1})_{i4}\Big]
\\
\displaystyle
\hspace{6cm}
\times\Big[z_1(\matQ^{-1})_{j1}+z_2(\matQ^{-1})_{j2}+z_3(\matQ^{-1})_{j3}+z_4(\matQ^{-1})_{j4}\Big]
\Bigg\},
\\
\hspace{14cm}
i\neq j
\end{cases}
\end{align}
}
The differentiations of \req{eq_skPsi2} with respect to $\hgss$, $\hqss$, $\hgssmm$ and $\hqssmm$ lead to
\begin{align}
&\gss = 1,
\\
\label{eq_skqss}
&\qss = 1 - 2D_{11},
\\
&\gssmm = 1,
\\
\label{eq_skqssmm}
&\qssmm = 1 - 2D_{44},
\end{align}
where $\qss=Q$ is known and one should extract $\hqss$ from the right hand side of \req{eq_skqss}. The factor 2 in Eqs.~(S\ref{eq_skqss}) and (S\ref{eq_skqssmm}) comes from the fact that $2\hqss$ and $2\qssmm$ are the elements of the matrix $\matQ$. The differentiations of \req{eq_skPsi2} with respect to $\hgssm$ and $\hqssm$ lead to
\begin{align}
&\gssm = \int\Dz\frac{1}{\cF(\vz, \mathbf{\hm}, \mathbf{\hq}, \mathbf{\hg})}\Bigg\{-2\cosh[z_1\!-\!z_4\!+\!\hms\!-\!\hmsm]e^{-(\hgssm\!-\!\hqssm)} + e^{(\hgssm\!-\!\hqssm)+\gamma}
\nonumber\\
&\quad\times\bigg[e^{-z_1-z_4-\hms-\hmsm}
\Omega_{-}(z_2, z_3, \hmh, \hme, \mathbf{\hq}, \mathbf{\hg})
+ e^{z_1+z_4+\hms+\hmsm}
\Omega_{+}(z_2, z_3, \hmh, \hme, \mathbf{\hq}, \mathbf{\hg})
\bigg]
\Bigg\},
\\
&\qssm = -D_{14} + \gssm.
\end{align}
From the definition  \req{eq_qssmab} of $\qssmab$ for $\alpha=\beta$, the size of self-sustained clusters is indeed given by $r=(1+\gssm)/2$. One can also differentiate \req{eq_skPsi2} with respect to $\gamma$ to show this relation. We then go on to differentiate with respect to $\hgsh$, $\hghsm$, $\hgse$ and $\hgesm$ by noting that
\begin{align}
\frac{\partial\Omega_{\pm}}{\partial\hgsh}=\frac{\partial\Omega_{\pm}}{\partial\hghsm}=\mp\frac{\partial\Omega_{\pm}}{\partial\hmh}
\\
\frac{\partial\Omega_{\pm}}{\partial\hgse}=\frac{\partial\Omega_{\pm}}{\partial\hgesm}=\mp\frac{\partial\Omega_{\pm}}{\partial\hme}
\end{align}
The differentiations of \req{eq_skPsi2} with respect to $\hgsh$, $\hqsh$, $\hghsm$ and $\hqhsm$ lead to
\begin{align}
&\gsh = \int\Dz\frac{1}{\cF(\vz, \mathbf{\hm}, \mathbf{\hq}, \mathbf{\hg})}\Bigg\{e^{(\hgssm\!-\!\hqssm)+\gamma}
\bigg[-e^{-z_1-z_4-\hms-\hmsm}
\frac{\partial\Omega_{-}}{\partial\hgsh}
+ e^{z_1+z_4+\hms+\hmsm}
\frac{\partial\Omega_{+}}{\partial\hgsh}
\bigg]
\Bigg\}
\\
&\qsh = -D_{12} + \gsh
\\
&\ghsm = \gsh
\\
&\qhsm = -D_{24} + \ghsm
\end{align}
Similarly, the differentiations with respect to $\hgse$, $\hqse$, $\hgesm$ and $\hqesm$ lead to
\begin{align}
&\gse = \int\Dz\frac{1}{\cF(\vz, \mathbf{\hm}, \mathbf{\hq}, \mathbf{\hg})}\Bigg\{e^{(\hgssm\!-\!\hqssm)+\gamma}
\bigg[-e^{-z_1-z_4-\hms-\hmsm}
\frac{\partial\Omega_{-}}{\partial\hgse}
+ e^{z_1+z_4+\hms+\hmsm}
\frac{\partial\Omega_{+}}{\partial\hgse}
\bigg]
\Bigg\}
\\
&\qse = -D_{13} + \gse
\\
&\gesm = \gse
\\
&\qesm = -D_{34} + \gesm
\end{align}

Finally, we differentiate  \req{eq_skPsi2} with respect to $\hghh, \hqhh, \hgee, \hqee, \hghe$ and $\hqhe$, which are elements of the covariance matrix $\matG$ of the Gaussian distribution in $\Omega_{\pm}$. We thus make use of the above lemma again and denote
{\small
\begin{align}
&D^{\Omega_{\pm}}_{ij}=& \nonumber \\ &\phantom{=}&\!\!\!\!\!\!\!\!\!\!\!\!\!\!\!\!\!\!\!\!\!\!\!\!
\begin{cases}
\displaystyle
\frac{1}{4\pi\sqrt{|\matG|}}\int \dx e^{-\frac{1}{2}\vx^T\matG^{-1}\vx}
\left\{-(\matG^{-1})_{ii}+\Big[z_1(\matG^{-1})_{i1}+z_2(\matG^{-1})_{i2}\Big]^2\right\}
\\
\displaystyle
\quad\times\Theta\Big[\Big(z_2\!+\!x_1\pm\!(\hgsh\!-\!\hqsh)\!\pm\!(\hghsm-\hqhsm) \!-\!\hmh\Big)\Big(z_3\!+\!x_2\!\pm\!(\hgse\!-\!\hqse)\!\pm\!(\hgesm-\hqesm) \!-\!\hme\Big)\Big].
\\
\hspace{14cm}
i=j
\\
\\
\displaystyle
\frac{1}{2\pi\sqrt{|\matG|}}\int \dx e^{-\frac{1}{2}\vx^T\matG^{-1}\vx}
\left\{-(\matG^{-1})_{ij}+\Big[z_1(\matG^{-1})_{i1}+z_2(\matG^{-1})_{i2}\Big]\Big[z_1(\matG^{-1})_{j1}+z_2(\matG^{-1})_{j2}\Big]\right\}
\\
\displaystyle
\quad\times\Theta\Big[\Big(z_2\!+\!x_1\pm\!(\hgsh\!-\!\hqsh)\!\pm\!(\hghsm-\hqhsm) \!-\!\hmh\Big)\Big(z_3\!+\!x_2\!\pm\!(\hgse\!-\!\hqse)\!\pm\!(\hgesm-\hqesm) \!-\!\hme\Big)\Big].
\\
\hspace{14cm}
i\neq j
\end{cases}
\end{align}
}
The differentiations of \req{eq_skPsi2} with respect to $\hghh, \hqhh, \hgee, \hqee, \hghe$ and $\hqhe$ lead to
\begin{align}
&\ghh = \int\Dz\frac{1}{\cF(\vz, \mathbf{\hm}, \mathbf{\hq}, \mathbf{\hg})}\Bigg\{ e^{(\hgssm\!-\!\hqssm)+\gamma}
\bigg[2e^{-z_1-z_4-\hms-\hmsm}
D^{\Omega_{-}}_{11}
+ 2e^{z_1+z_4+\hms+\hmsm}
D^{\Omega_{+}}_{11}\bigg]
\Bigg\}
\\
&\qhh = -2D_{22} + \ghh
\\
&\gee = \int\Dz\frac{1}{\cF(\vz, \mathbf{\hm}, \mathbf{\hq}, \mathbf{\hg})}\Bigg\{ e^{(\hgssm\!-\!\hqssm)+\gamma}
\bigg[2e^{-z_1-z_4-\hms-\hmsm}
D^{\Omega_{-}}_{22}
+ 2e^{z_1+z_4+\hms+\hmsm}
D^{\Omega_{+}}_{22}\bigg]
\\
&\qee = -2D_{33} + \gee
\\
&\ghe = \int\Dz\frac{1}{\cF(\vz, \mathbf{\hm}, \mathbf{\hq}, \mathbf{\hg})}\Bigg\{ e^{(\hgssm\!-\!\hqssm)+\gamma}
\bigg[e^{-z_1-z_4-\hms-\hmsm}
D^{\Omega_{-}}_{12}
+ e^{z_1+z_4+\hms+\hmsm}
D^{\Omega_{+}}_{12}\bigg]
\\
&\qhe = -D_{23} + \ghe
\end{align}
We iterate all the above equations numerically to obtain the solution of all the unknown variables.

\subsection{Final expression of $\ln\skZ$}

Finally, we use the relations derived in the previous subsection to express $\ln\skZ$ in terms of $M$, $Q$, $\hms$, $\hqss$ and all the other variables of $m$ and $q$, which lead to
{\small
\begin{align}
\label{eq_lnskZ}
\frac{1}{N}\ln\skZ
&= -M\hms - J_0\msm\me + (Q-1)\hqss +\frac{J^2}{2}(\gee-\qee)
\nonumber\\
&+\frac{J^2}{2}\Big(\left(\qsh\right)^2 + 2\qhe\qssm + 2\qhsm\qse + \qee\qssmm + \left(\qesm\right)^2 \Big)
\nonumber\\
&-\frac{J^2}{2}\Big(\left(\gsh\right)^2 + 2\ghe\gssm + 2\ghsm\gse + \gee\gssmm + \left(\gesm\right)^2 \Big)
\nonumber\\
& + \int\Dz\log
\Bigg\{2\cosh[z_1\!-\!z_4\!+\!\hms\!-\!J_0\me]e^{-J^2(\ghe\!-\!\qhe)} + e^{J^2(\ghe\!-\!\qhe)\!+\!\gamma}
\nonumber\\
&\times\bigg[e^{-z_1-z_4-\hms-J_0\me}
\Omega_{-}(z_2, z_3, M, \msm, \mathbf{q}, \mathbf{\qzero})
+ e^{z_1+z_4+\hms+J_0\me}
\Omega_{+}(z_2, z_3, M, \msm, \mathbf{q}, \mathbf{\qzero})
\bigg]\Bigg\}
\end{align}
}
where $\int \Dz$ represents the Gaussian integration
\begin{align}
\frac{1}{4\pi^2\sqrt{|\matQ|}}\int \dz e^{-\frac{1}{2}\vz^T\matQ^{-1}\vz},
\nonumber
\end{align}
with the covariance matrix $\matQ$ given by
\begin{align}
\matQ = J^2
\begin{pmatrix}
2\hqss/J^2 & \qsh & \qhsm & \qhe
\\
\qsh & Q & \qssm & \qse
\\
\qhsm & \qssm & \qssmm & \qesm
\\
\qhe & \qse & \qesm & \qee
\end{pmatrix}.
\end{align}
The functions $\Omega_{\pm}$ are given by
\begin{align}
\label{eq_omega2}
&\Omega_\pm(z_2, z_3, M, \msm, \mathbf{q}, \mathbf{\qzero})
\nonumber\\
& = \frac{1}{2\pi\sqrt{|\matG|}}\int \dx e^{-\frac{1}{2}\vx^T\matG^{-1}\vx}\Theta\Big[\Big(z_2\!+\!x_1\pm\!J^2(\gsh\!-\!\qsh)\!\pm\!J^2(\gse-\qse) \!-\!J_0M\Big)
\nonumber\\
&\hspace{6cm}
\times\Big(z_3\!+\!x_2\!\pm\!J^2(\ghsm\!-\!\qhsm)\!\pm\!J^2(\gesm-\qesm) \!-\!J_0\msm\Big)\Big].
\end{align}
where the covariance matrix $\matG$ is given by
\begin{align}
\matG = J^2
\begin{pmatrix}
1-Q & \gssm-\qssm
\\
\gssm-\qssm & 1-\qssmm
\end{pmatrix}.
\end{align}
In the limit of $J_0\to \infty$, which reduces to the Ising model case, one can show that: (i) $\Omega_{\pm}\to\Theta(M\msm)$ when $M\neq 0$ and $\msm\neq 0$; (ii) with the ansatz $\hms\propto J_0$, $\cosh(z_1-z_4+\hms-J_0\me)\approx \cosh(\hms-J_0\me)$ and $e^{\pm(z_1+z_4+\hms+J_0\me)}\approx e^{\pm(\hms+J_0\me)}$, which implies $\ln \skZ$ is independent of the matrix $\matQ$ and each $\qzero$ and its corresponding $q$ variable will become equal; (iii) these lead to $\hms=\beta J_0M$ which implies (ii) is a self consistent assumption. These results show that in the limit of $J_0\to\infty$, $\skZ$ in \req{eq_lnskZ} reduces to $\isingZ$ in \req{eq_isingZ_solution} of the Ising model.

\subsection{Derivation of $\ln B(M,Q)$}

We will derive an expression for the spin entropic contribution $\ln B(\{\Ma\},\{\Qab\})$ given by
\begin{align}
&\ln B(\{\Ma\},\{\Qab\}) = \lim_{n\to 0}\frac{1}{n}
\Bigg[\trace_{\{\sia\}}
\prod_{\alpha}\delta\left(\frac{\sum_i\sia}{N}\!-\!\Ma\right)\!
\prod_{\alpha\neq\beta}\delta\left(\frac{\sum_i\sia\sib}{N}\!-\!\Qab\right)-1\Bigg].
\end{align}
We start with
\begin{align}
&\trace_{\{\sia\}}\prod_{\alpha}\delta\left(\frac{\sum_i\sia}{N}\!-\!\Ma\right)\!
\prod_{\alpha\neq\beta}\delta\left(\frac{\sum_i\sia\sib}{N}\!-\!\Qab\right)
\nonumber\\
&\qquad=\trace_{\{\sia\}}
\prod_{\alpha}\left[\int \frac{d \hMa}{2\pi}e^{i \Ma\hMa - i\hMa\frac{\sum_i\sia}{N}}\right]
\prod_{\alpha\beta}\left[\int \frac{d \hQab}{2\pi}e^{i \Qab\hQab - i\hQab\frac{\sum_i\sia\sib}{N}}\right].
\nonumber\\
&\qquad\propto\prod_{\alpha}\int \frac{d \hMa}{2\pi}\prod_{\alpha\beta}\int \frac{d \hQab}{2\pi}e^{Nn\BPsi}
\end{align}
where we have used the change of variables $\hMa\!\to\!iN\hMa$ and $\hQab\to iN\hQab$ to arrive at the last line, with $n\BPsi$ given by
\begin{align}
n\BPsi = -\suma \Ma\hMa -\sum_{\alpha, \beta}\Qab\hQab + \log\left\{\trace_{\{\sa\}} \exp\left[\suma\hMa\sa+\sum_{\alpha, \beta}\hQab\sa\sbb\right]\right\}.
\end{align}
Using the replica symmetric ansatz $\Ma=M$ and $\hMa=\hM$ for all $\alpha$, $\Qab=Q$ and $\hQab=\hQ$ for all $\alpha\neq\beta$, and $\Qab=1$ and $\hQab=\hat{C}$ for all $\alpha=\beta$, we arrive at
{\small
\begin{align}
n\BPsi &= -n M\hM - n(n-1)Q\hQ - n\hat{C} + \log\left\{\trace_{\{\sa\}} \exp\left[\hM\suma\sa+\hQ\left(\suma\sa\right)^2+n(\hat{C}-\hQ)\right]\right\}
\nonumber\\
&=-n M\hM - n(n-1)Q\hQ - n\hQ +\log\left\{\frac{1}{\sqrt{2\pi}}\int dz e^{-\frac{z^2}{2}}\prod_{\alpha}\left[\sum_{s=\pm 1}e^{\sqrt{2\hQ}\sa+\hM\sa}\right]\right\}.
\end{align}
}
In the limit of $n\to 0$, $\BPsi$ is given by
\begin{align}
\label{eq_BPsi}
\lim_{n\to 0}\BPsi = -M\hM +Q\hQ - \hQ +\frac{1}{\sqrt{2\pi}}\int dz e^{-\frac{z^2}{2}}\log\left[2\cosh\left(\sqrt{2\hQ}z+\hM\right)\right].
\end{align}
By the method of steepest descent, we differentiate \req{eq_BPsi} with respect to $\hM$ and $\hQ$ to obtain
\begin{align}
M &= \frac{1}{\sqrt{2\pi}}\int dz e^{-\frac{z^2}{2}}\tanh\left(\sqrt{2\hQ}z+\hM\right),
\\
Q &= 1- \frac{1}{\sqrt{2\pi}}\int dz e^{-\frac{z^2}{2}}\tanh\left(\sqrt{2\hQ}z+\hM\right)\frac{z}{\sqrt{2\hQ}}
\nonumber\\
&=  \frac{1}{\sqrt{2\pi}}\int dz e^{-\frac{z^2}{2}}\tanh^2\left(\sqrt{2\hQ}z+\hM\right)
\end{align}
which are identical to the equation of states in the original SK model when the unknown $\hM=\beta J_0M$ and $\hQ=\beta^2 J^2 Q/2$. Substituting this solution into \req{eq_BPsi}, we obtain an expression for $\ln B(M,Q)$, given by
\begin{align}
\ln B(M,Q) &= \lim_{n\to 0}\BPsi
\nonumber\\
&=\!-\beta J_0 M^2\!+\!\frac{\beta^2 J^2}{2}Q(Q\!-\!1)\!+\!\frac{1}{\sqrt{2\pi}}\int dz e^{-\frac{z^2}{2}}\log\left[2\cosh\left(\beta J\sqrt{Q}z+\beta J_0 M\right)\right].
\end{align}
